\documentstyle[aps,prb
,twocolumn
,floats,epsf,rotate
]{revtex}

\def\divisor{\,|\,}              
\def\notdiv{\,|\mkern-7mu/\,}    

\catcode`@=11  
\def\@oddfoot{\mbox{}\hfil\rm \thepage{}
\footnotesize(November 30, 1997)\hfil\mbox{}}
\catcode`@=12 

\begin{document}

\title{
   Toward semiclassical theory of
   quantum level correlations
   of generic  chaotic systems. }
\author{ Daniel L. Miller }
\address{
 Dept. of Physics of Complex Systems,\\
 The Weizmann Institute of science,
 Rehovot, 76100 Israel                \\
 e-mail  fndaniil@wicc.weizmann.ac.il
}
\date{November 30, 1997}
\twocolumn[
\maketitle\begin{abstract}
   \widetext\hsize                  
   \columnwidth\leftskip=0.10753    
   \textwidth\rightskip\leftskip    
   \nointerlineskip\small\relax     
		  In the present work we study the two-point correlation function
		  $R(\varepsilon)$ of the quantum mechanical spectrum of a classically
		  chaotic system. Recently this quantity has been computed for chaotic
		  and for disordered systems using periodic orbit theory and field
		  theory. In this work we present an independent derivation, which is
		  based on periodic orbit theory. The main ingredient in our approach
		  is the use of the spectral zeta function and its autocorrelation
		  function $C(\varepsilon)$. The relation between $R(\varepsilon)$ and
		  $C(\varepsilon)$ is constructed by making use of probabilistic
		  reasoning similar to that which has been used for the derivation of
		  the Hardy -- Littlewood conjecture. We then convert the symmetry
		  properties of the function $C(\varepsilon)$ into relations between
		  the so-called diagonal and  the  off-diagonal parts of
		  $R(\varepsilon)$. Our results are valid for generic systems with
		  broken time reversal  symmetry, and with non-commensurable periods of
		  the periodic orbits.  
   \\ \\
\end{abstract}
]
\narrowtext


\section{Introduction}
\label{sec:intr}

\begin{figure*}
\unitlength=1mm
\begin{picture}(178,135)
\put(0,0){\epsfxsize=178mm\epsffile{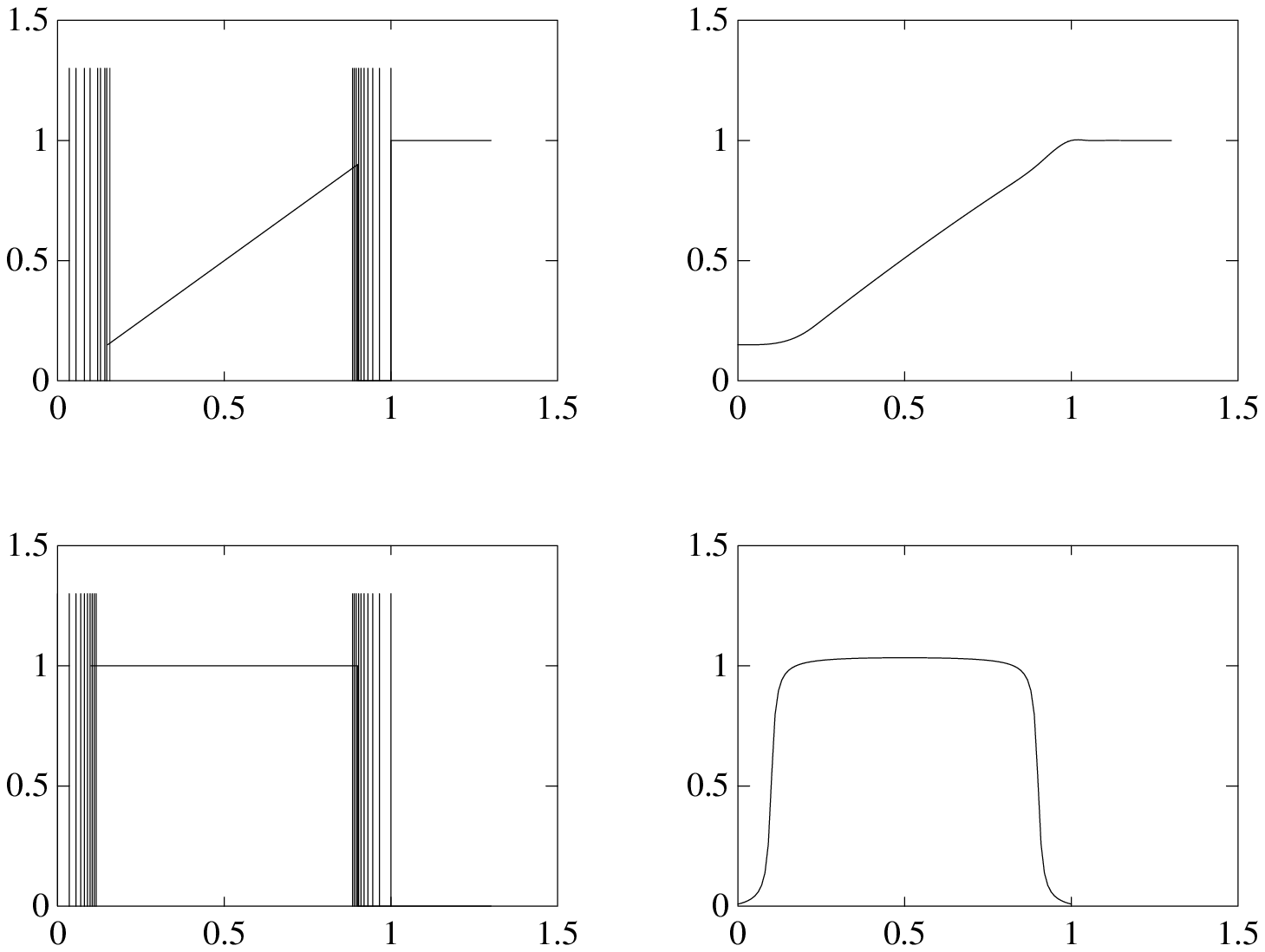}}
{\Large
\put(3,30){\rotate[l]{$K^\zeta(\tau)$}}
\put(3,105){\rotate[l]{$K(\tau)$}}
\put(90,30){\rotate[l]{$K^\zeta(\tau)$}}
\put(90,105){\rotate[l]{$K(\tau)$}}
\put(45,3){$\tau$}
\put(135,3){$\tau$}
\put(45,73){$\tau$}
\put(135,73){$\tau$}
\put(45,125){(a)}
\put(135,125){(b)}
\put(45,55){(c)}
\put(135,55){(d)}
\put(22.5,80.){\makebox(0,0)[cc]{$\tau_{\text{erg}}$}}
\put(110.,80.){\makebox(0,0)[cc]{$\tau_{\text{erg}}$}}
\put(22.5,11.){\makebox(0,0)[cc]{$\tau_{\text{erg}}$}}
\put(110.,11.){\makebox(0,0)[cc]{$\tau_{\text{erg}}$}}
\put(62.0,79.5){\makebox(0,0)[lc]{$=\tau_H$}}
\put(151.,79.5){\makebox(0,0)[lc]{$=\tau_H$}}
\put(62.0,10.5){\makebox(0,0)[lc]{$=\tau_H$}}
\put(151.,10.5){\makebox(0,0)[lc]{$=\tau_H$}}
}
\end{picture}
\caption{Form-factors derived from correlation functions of density of states
and spectral determinant. Diagrams (a), and (c) correspond to a small chaotic
system, where one can observe fingerprints of short periodic orbits in the
form-factors, shown as $\delta$-functional peaks. Diagrams (b) and (d)
correspond to diffusive systems, which can be considered as huge complex
chaotic systems. One can observe smoothing of the form-factors near $\tau=0$
and $\tau=\tau_H$. Particular shape of this smoothing contains system
specific information like dimensionality and characteristics of disorder. }
\label{fig:allff} \end{figure*}

Quantum chaology\cite{Berry-87} has attracted the attention of the physics
community after the discovery\cite{Bohigas-84} that the spectral
correlations  of classically chaotic systems are universal. They exhibit a
strong level repulsion, which induces a non-trivial two-point correlation
function  $R(\varepsilon)$ of the density of states.

The spectral rigidity of a generic chaotic system was computed for the first
time by Berry\cite{Berry-feb85}. He expanded the density of states over
periodic orbits by making use of Gutzwiller's trace
formula\cite{Gutzwiller-book91}. He estimated number of orbits of a given
length from the Hannay -- Ozorio De Almeida sum
rule\cite{Hannay-OzorioDeAlmeida} and computed the so-called diagonal part of
the form-factor $K_{\text{diag}}(\tau)$. The form-factor $K(\tau)$ is a
function of time, and is the Fourier transform of the two-point
correlation function $R(\varepsilon)$.

The semiclassical theory of the form-factor distinguishes between three main
time scales, see diagram Fig.~\ref{fig:allff}(a). The contribution of the
short periodic orbits is shown schematically as a sequence of $\delta$-peaks,
and the relevant time scale is the period of the shortest periodic orbit
$\tau_p$. This non-universal behavior of the form-factor prevails between
$\tau=0$ and the ergodic time $\tau_{\text{erg}}$.\cite{Ergodic-time-note} 
Berry found that $K(\tau) \propto \tau$ for $\tau_H \gtrsim \tau \gtrsim
\tau_{\text{erg}}$, and showed that this explains the level repulsion.

In the present work we are primarily interested in the semiclassical
(periodic-orbit) theory of the form-factor for $\tau\gtrsim\tau_H$, where
$\tau_H$ is the Heisenberg time. This is the third time scale. It is of
quantum mechanical nature, because the Heisenberg time is proportional to the
mean density of states. In Fig.~\ref{fig:allff} we have chosen units of time
such that $\tau_H=1$. Due to the discrete nature of the density of states,
the form-factor becomes constant for times much larger than $\tau_H$, see
Fig.~\ref{fig:allff}(a).  The semiclassical theory for the form-factor for
$\tau\gtrsim\tau_H$ has been analyzed recently  by Bogomolny and
Keating.\cite{Bogomolny-aug96} They found the fingerprints of the short
periodic orbits in the vicinity of the Heisenberg time as shown in
Fig.~\ref{fig:allff}(a) by $\delta$-peaks near $\tau=\tau_H$.

The most difficult problem addressed by Bogomolny and Keating was the
computation of the off-diagonal part of the two-point correlation function
$R_{\text{off}} (\varepsilon)$. Argaman {\it et al}\cite{Argaman-dec93}
pointed out that  it is possible to compute $R_{\text{off}}(\varepsilon)$ if
the two-point statistics of the classical actions is known. Let us call  the
set of actions of all primitive periodic orbits (PPOs) the {\em length
spectrum}. The density of states is related to the length spectrum by the
Gutzwiller trace formula\cite{Gutzwiller-book91}. Therefore correlations of
eigenenergies should be related to correlations of actions.

Our approach to the semiclassical evaluation of $R_{\text{off}}
(\varepsilon)$ further develops the approach of Argaman
{\it et al}\cite{Argaman-dec93} and their
followers\cite{Doron-prl97,Cohen-Primack-Smilansky} but makes use of a
different starting point. The first new element is the study of the spectral
zeta function $\zeta(E)$, its autocorrelation function $C(\varepsilon)$, and
its Fourier transform $K^\zeta(\tau)$.

The spectral zeta function $\zeta(E)$ is an important tool for an analytic or
numeric computation of the energy levels of any quantum mechanical system.
The zeros of this function are the eigenenergies of the Hamiltonian. A
unique definition of $\zeta(E)$ is to be given later.  Voros\cite{Voros-88}
has proposed to compute this function by making use of a product over the
periodic orbits. Berry and Keating\cite{Berry-Keating-jun90} have expanded
this product over the composite periodic orbits (CPOs), see precise formula
below. The spectral zeta function is a smooth function that has no
$\delta$-peaks like the density of states does. Therefore it is a good idea to
characterize level statistics of chaotic systems by the $\zeta(E)$
autocorrelation function.  This was done by Kettemann, Klakow and Smilansky,
who also computed this autocorrelation function $C(\varepsilon)$ for two and
three dimensional Sinai billiards.\cite{Smilansky-sct97}

The diagram Fig.~\ref{fig:allff}(c) schematically shows $K^\zeta(\tau)$. For
systems with broken time reversal symmetry, as in Fig.~\ref{fig:allff}, one
can see the separation of the time scales very clearly.  The short time
behavior of $K^\zeta(\tau)$ is determined by the short CPOs, and it is
represented by $\delta$-peaks.  They become dense giving a constant behavior
of $K^\zeta(\tau)$ for $\tau>\tau_{\text{erg}}$, see
Ref.\onlinecite{Smilansky-sct97}. It is known that the spectral zeta function
has to satisfy a functional equation.\cite{Keating-92} This equation can be
derived from the definition of $\zeta(E)$ and implies an exact mirror
symmetry of $K^\zeta(\tau)$ around the half Heisenberg time:
$K^\zeta(\tau_H/2 + \tau)=K^\zeta(\tau_H/2 - \tau)$. For this reason one has
to observe the fingerprints of the short composite periodic orbits near
$\tau_H$; they are shown as $\delta$-peaks near $\tau_H$ in
Fig.~\ref{fig:allff}(c). Therefore, if the short time behavior of
$K^\zeta(\tau)$ is known, then the behavior of $K^\zeta(\tau)$ near $\tau_H$
is also known.

Similar to the case of $R(\varepsilon)$, the periodic orbits expansion of
$\zeta(E)$ allows one to separate $C(\varepsilon)$ and $K^\zeta(\tau)$ into
diagonal and off-diagonal parts.  Following Berry\cite{Berry-feb85}, we can
assume that the off-diagonal part of $K^\zeta(\tau)$ vanishes for $\tau <
\tau_H/2$. This assumption together with the mirror symmetry of
$K^\zeta(\tau)$ implies an explicit connection between the diagonal and the
off-diagonal parts of $K^\zeta(\tau)$.  Physically, this relation has to be
interpreted as a {\em quantum-classical time scale separation}.

Since $\zeta(E)$ is expressed in terms of CPOs by the Berry -- Keating
formula\cite{Berry-Keating-jun90}, the autocorrelation function
$C(\varepsilon)$  is related to correlations in the {\em composite length
spectrum}, which is the set of all CPOs. In this way we introduce an analog
of length correlations, which were used by Argaman {\it et
al}\cite{Argaman-dec93} to discuss the two-point spectral functions. One of
the central ideas of the present work is to compute the correlations in the
composite length spectrum by making use of the mirror symmetry of
$K^\zeta(\tau)$. More over, the time scale separation provides an explicit
dependence of $K^\zeta(\tau)$ on $\tau_H$. The two point correlation function
of composite actions is just the Fourier transform of $K^\zeta(\tau)$ with
respect to $1/\hbar$ hidden in $\tau_H$. This idea is similar to one which
has been suggested by Balian and Bloch.\cite{Balian-Bloch-jan-74}

The length or the action of the given composite periodic orbit is simply the
algebraic sum of the lengths or the actions of the PPOs forming this
composite orbit.  For this reason, action correlations of CPOs have to be
related to action correlations of PPOs. In the present work we construct an
integral relation between these correlation functions by making use of simple
probabilistic arguments. This relation is generic and independent of  the
dimensionality of the system, however it might be dependent on the symmetry
of the system.  We perform calculations only for systems which does not have
spatial symmetries, because we assume that the composite length spectrum
is non-degenerate.

The statistical relation between correlations of CPOs and correlations of PPOs
can be converted by the inverse Fourier transform and a special regularization
procedure to a relation between the off-diagonal parts of $K(\tau)$ and
$K^\zeta(\tau)$. Calculations show that the behavior of $K(\tau)$
near $\tau_H$, which is shown schematically in Fig.~\ref{fig:allff}(a),
reproduces the behavior of $K^\zeta(\tau)$ near $\tau_H$, see
Fig.~\ref{fig:allff}(c).

The ideas of this paper are best understood by the following logical flow:
\begin{enumerate}
\item
   The off-diagonal part of $K(\tau)$ is the Fourier transform with respect to
   $1/\hbar$ of the two-point statistics of actions of PPOs.
\item
   The two-point statistics of actions of PPOs can be computed from the
   two-point statistics of actions of CPOs.
\item
   The two-point statistics of actions of CPOs is determined by the
   off-diagonal part of $K^\zeta(\tau)$.
\item
   The diagonal and off-diagonal parts of $K^\zeta(\tau)$ are connected,
   because $\zeta(E)$ satisfies the functional equation.
\item
   For the systems with broken time reversal symmetry there is a clear
   separation between the classical scale (time of mixing) and the quantum
   mechanical scale (time of quantum recurrence). In this case we can compute
   the off-diagonal part of $K^\zeta(\tau)$ explicitly, if the diagonal part is
   known.
\item
   The diagonal parts of either $K(\tau)$ and $K^\zeta(\tau)$ or
   $R(\varepsilon)$ and $C(\varepsilon)$ have well known periodic orbit
   expansions, which can be evaluated both numerically and by making use of
   sum rules.\cite{Berry-feb85,Smilansky-sct97}
\end{enumerate}

The two-point statistics of actions is a property of classical dynamics and
it is independent of $\hbar$.  Unfortunately, there are only a few examples
of chaotic systems where a sufficient number of  periodic orbits is
known and the action correlations can be computed directly from the classical
Hamiltonian.\cite{Cohen-Primack-Smilansky} We circumvent this difficulty by
computing action correlations from another source, namely from the diagonal
part of $C(\varepsilon)$, which is also a purely classical
object.\cite{Smilansky-sct97}

The integral relation connecting the correlations of the PPOs and the CPOs can
be applied directly to derive the well known correlations between prime
numbers. As we show in Appendix~\ref{sec:RZexample}, the Hardy -- Littlewood
expression\cite{Hardy-Littlewood-22} can be reproduced in this framework. 
We emphasize this application as a very stringent test for our probabilistic
method.

Large chaotic systems, for example billiards with a large number of
scatterers or disordered systems, usually have $\tau_{\text{erg}} \gg
\tau_p$, since $\tau_{\text{erg}}$ is the time of a diffusion through the
system. The system specific features of $K(\tau)$ and $K^\zeta(\tau)$ are
smooth, see Figs.~\ref{fig:allff}(b) and \ref{fig:allff}(d). The condition
$\tau_{\text{erg}}\gg \tau_p$ allows us to express the diagonal part of
$C(\varepsilon)$  as the Fredholm determinant of the diffusion propagator.  In
this way our expressions for $R(\varepsilon)$ reproduce precisely the field
theory result of Andreev and Altshuler.\cite{Andreev-Altshuler-95}

At this point it is appropriate to review the important work on the level
statistics of disordered systems which had a very important impact on the
development of the semiclassical theories presented here and
elsewhere.\cite{Bogomolny-aug96,Cohen-Primack-Smilansky} Field theory was
used by Efetov to compute the level statistics of small metallic samples, see
the review paper\cite{Efetov-jan83}. He considered the electron moving in a
random potential of impurities and confined by sample boundaries. He assumed
that the fluctuations of the fields are uniform across the sample and
obtained universal results for the level statistics.

Altshuler and Shklovskii\cite{Altshuler-jul86} made use of perturbation
theory and expressed the so-called perturbative part of $R(\varepsilon)$ in
terms of the density-density correlation function, which is the propagator of
the diffusion equation. Argaman {\em et al}\cite{Argaman-feb93} showed that
the perturbative part of $R(\varepsilon)$ is nothing but the diagonal part of
$R(\varepsilon)$ mentioned before.

The diagonal (or perturbative) part of $R(\varepsilon)$ is singular at small
energies; the only way to remove this singularity is to compute the
off-diagonal (or non-perturbative) part of $R(\varepsilon)$. Andreev and
Altshuler\cite{Andreev-Altshuler-95} computed this term by making use of
non-perturbative field theory and obtained the answer in terms of the same
density-density correlation function. These authors repeated Efetov's
calculations, but allowed the spatial fluctuations of the superfields.

The free energy functional in Efetov's theory describes
the diffusion modes. These modes have to satisfy the diffusion equation. This theory
was remarkably generalized by Muzykantskii and Khmelnitskii\cite{MK-jul95}.
They obtained the free energy functional describing the eigenmodes of the
kinetic equation. Such solutions to the kinetic equation have a meaning of
the density-density correlations\cite{landau-corrfun}.

Muzykantskii and Khmelnitskii also suggested that the modes of the kinetic
equation should be replaced by the modes of the Liouvillian operator if one
goes from diffusive to chaotic systems. Agam{\em et al}\cite{AAA-dec95} and
Andreev {\em et al}\cite{ASAA-96} constructed field theory for chaotic
systems by averaging over the energy and obtained the Liouvillian operator in
the kinetic part of the free energy functional.  Both types of field theory
show that the spectral statistics of a chaotic system is described  by the
determinant of the Liouvillian operator.

The connection between the diagonal part of the correlation function and the
determinant of the Liouvillian operator is almost trivial in the framework of
periodic orbit theory. This connection is different from the field theory
prediction due to the terms containing repetitions of the PPOs. The periodic
orbit theory expression for the off-diagonal part of the correlation function
obtained by Bogomolny and Keating\cite{Bogomolny-aug96} is also different
from the field theory result\cite{ASAA-96}. Therefore, the derivation of the
Andreev -- Altshuler result in terms of the action correlations is interesting,
and it also gives additional information about the autocorrelation function
of the spectral determinant. The latter has been obtained by using random
matrix theory\cite{Andreev-Simons-sep95,Haake-jan96} and random
polynomial theory\cite{Bogomolny-dec96}, as opposed to field theory.

All correlation functions employed in the present work are defined in
Secs.~\ref{sec:definitions} and \ref{sec:defZfun}. In Sec.~\ref{sec:pnt} very
simple probabilistic arguments will help us  to build the integral equation
connecting the correlations of CPOs with the correlations of PPOs. The
behavior of form-factors near the Heisenberg time is considered in
Sec.~\ref{sec:uff}. In the same section our results are compared with the
universal random matrix theory predictions and the results of
Refs.\onlinecite{Andreev-Altshuler-95,Bogomolny-aug96}.  We discuss the
physics of the action correlations in Sec.~\ref{sec:disc}. We also summarize
our results in Sec.~\ref{sec:disc}.


\section{Definitions of objects related to the spectral correlations.}
\label{sec:definitions}

We start to build our theory for the specific example of the chaotic
billiard. Let us also put the Aharonov -- Bohm flux through the billiard in
order to break the time reversal symmetry. Let us also assume that the system
has no spatial symmetries. The classical motion of a charged particle in the
billiard is finite and therefore one can associate two sets with this
system.  The first set, $\{l_p\}$, is formed by the lengths of all the 
PPOs\cite{Smilansky-rev95}, and each orbit is labeled by the index $p$. Each
orbit makes $\mu_p$ windings around the Aharonov -- Bohm flux. We distinguish
the orbits with positive and negative winding numbers, i.e. the paths
going clockwise and counterclockwise around the flux tube. Therefore, the
length spectrum $\{l_p\}$ is degenerate, the lengths of the orbits with 
non-zero winding numbers appear twice.

The second set associated with the billiard is the quantum mechanical
spectrum. It is formed by the wave vector magnitudes $\{k_n\}$.  Each
of them corresponds to the eigenvalue of the Hamiltonian $E_n = E(k_n)$ and
$E(k)=\hbar^2k^2/(2m)$, where $m$ is the mass of a particle.

\begin{mathletters}
The density of states is a sequence of $\delta$-peaks, and it is
usually computed by making use of the Green function of the system:
\begin{equation}
   d(k) \equiv \sum_{n=1}^\infty \delta(k-k_n)
   = - {1\over \pi} \text{Im}\,\text{Tr}\,\hat G^r(k) E'(k)\;.
\label{eq:dfn.1a}
\end{equation}
Here $\hat G^r(k)$ is the retarded Green function of the system, and the
prime means derivative with respect to $k$. The trace of $\hat G^r(k)$ can be
represented as a sum over the periodic orbits. Such an expansion is called a
trace formula.\cite{Gutzwiller-book91} The derivation of this expansion for a
billiard can be found in the review Ref.\onlinecite{Smilansky-rev95}. The
result is
\begin{equation}
  \text{Tr}\hat G^r(k)E'(k) =
  - \sum_{p}
    \sum_{r=1}^\infty
  {i l_p e^{ik rl_p-i\nu_p \pi/2 + i\phi\mu_p}
    \over e^{\lambda_p r l_p /2} - e^{-\lambda_p r l_p /2}
  } - i \pi \bar d(k)\;,
\label{eq:dfn.1b}
\end{equation}
where the index $p$ runs over the PPOs, the length of each orbit is $l_p$,
the Lyapunov exponent of each orbit is $\lambda_p l_p$, and the Maslov index is
$\nu_p$. We also broke the time reversal symmetry by adding the phase due to
Aharonov -- Bohm flux $\phi$ multiplied by the winding number of the trajectory
$\mu_p$. Lyapunov exponents in Eq.~(\ref{eq:dfn.1b}) are defined per unit
length of a trajectory and not per number of scatterings as in
Ref.\onlinecite{Smilansky-rev95}. This formula also contains the smooth part
of the density of states $\bar d(k)$, which is proportional to the volume of
the system. Equations~(\ref{eq:dfn.1a}) and (\ref{eq:dfn.1b}) show that the
density of states contains an oscillating part. It is defined as
\begin{eqnarray}
   d_{\text{osc}}(k) \equiv d(k) - \bar d(k)\;,
\label{eq:dfn.1}
\end{eqnarray}
and is proportional to the imaginary part of the first term on the right
hand side of Eq.~(\ref{eq:dfn.1b}).
\label{eq:dfn.2}
\end{mathletters}


The correlation function of the quantum mechanical spectrum and its Fourier
transform have to be defined in a special way, because of the
$\delta$-functional form of the density of states and the discrete form of
the expansion Eq.~(\ref{eq:dfn.1b}).  We define the correlation function by
making use of the averaging over $k$ with Gaussian weight
\begin{eqnarray}
  R(\varepsilon,k)
  &\equiv&
  e^{-\Delta l^2\varepsilon^2/2}
  \,
  \int_{-\infty}^\infty
  {
     dq   \over
     \sqrt{ 2 \pi }\Delta k
  }
  \,
  e^{- {(q-k)^2 \over 2\Delta k^2}}
\nonumber\\
  &\times&
  d_{\text{osc}}(q+{\varepsilon\over 2})
  d_{\text{osc}}(q-{\varepsilon\over 2})
\label{eq:dfn.3}
\end{eqnarray}
and this definition of the correlation function is valid for $\varepsilon\ll k$.
We emphasize that $\varepsilon$ in our definition is the difference of the
wave numbers and not of the energies as accepted.

\begin{figure}
\unitlength=1mm
\begin{picture}(88,72)
\put(00.,00.){\epsfxsize=100mm\epsffile{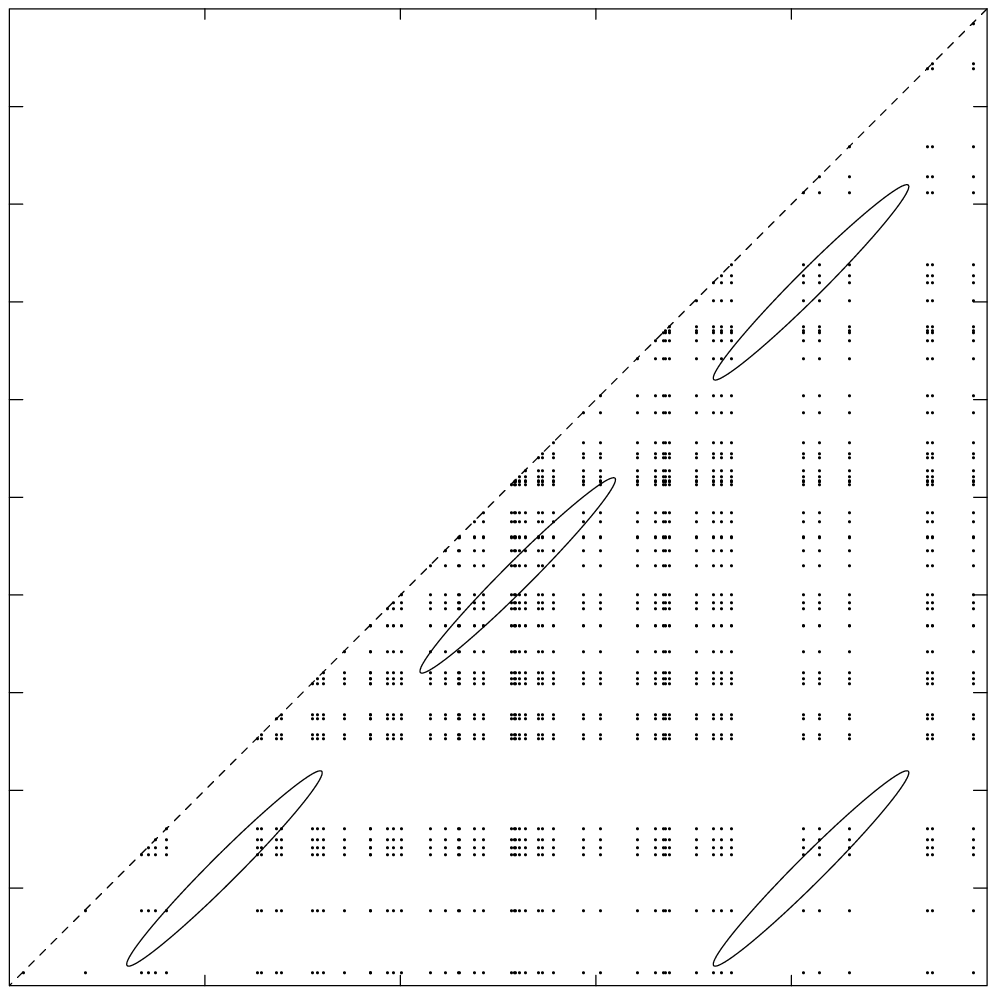}}
\put(22.00,10.00){\line(1,1){62.00}}
\put(41.00,37.00){\vector(1,-1){4.00}}
\put(52.00,26.00){\vector(-1,1){4.00}}
\put(41.00,37.00){\line(1,-1){14.00}}
\put(53.00,23.00){\makebox(0,0)[cc]{$\varepsilon$}}
\put(45.00,27.00){\line(1,1){6.00}}
\put(38.00,32.00){\line(1,1){17.00}}
\put(39.00,33.00){\vector(1,1){3.00}}
\put(58.00,52.00){\vector(-1,-1){4.00}}
\put(53.00,49.00){\line(1,-1){10.00}}
\put(50.00,45.00){\makebox(0,0)[rb]{$\sim \Delta k$}}
\put(55.00,2.00){\makebox(0,0)[cb]{$k_n$}}
\put(7.00,38.00){\makebox(0,0)[cc]{$k_{n'}$}}
\put(21.70,38.00){\line(1,0){3.00}}
\put(20.00,38.00){\makebox(0,0)[rc]{$k^{(b)}$}}
\put(21.70,20.00){\line(1,0){3.00}}
\put(20.00,20.00){\makebox(0,0)[rc]{$k^{(a)}$}}
\put(21.70,56.00){\line(1,0){3.00}}
\put(20.00,56.00){\makebox(0,0)[rc]{$k^{(c)}$}}
\end{picture}
\caption{Schematic diagram demonstrating how to compute the correlation
function of  energy levels. Each point represents pair of the energy levels
$(k_{n},k_{n'})$.  The correlation function $R(\varepsilon, k)$ picks up the
points inside one of the ellipses shown in the diagram. Three ellipses near
the diagonal are drawn for $k$ equal to $k^{(a)}$, $k^{(b)}$, $k^{(c)}$, and
$\varepsilon \ll \Delta k$. We draw the fourth ellipse for the case
$\varepsilon  \sim \Delta k$ and the choice of relevant $k$ becomes
ambiguous.  
} \label{fig:dfn.pol}\end{figure}

The definition of the correlation function Eq.~(\ref{eq:dfn.3}) contains the
product of two $\delta$-functions from two densities of states and only one
averaging.  Practically, the $\delta$-function remaining after the
integration in Eq.~(\ref{eq:dfn.3}) has to be replaced by a smoothed function
of the width smaller than the mean level spacing.  Then, the definition
Eq.~(\ref{eq:dfn.3}) can be understood geometrically. The two-point
correlation function is proportional to the difference between the number of
the level pairs $(k_{n},k_{n'})$ inside the region, which we showed
schematically as an ellipse in Fig.~\ref{fig:dfn.pol}, and the square of the 
mean density of states  multiplied by the area of this region. The ``length''
of the ellipse is of the order of $\Delta k$ and the ``width'' has to be
smaller than  the mean level spacing.

We show three ellipses demonstrating the correlation function computed for
$k$ equal to $k^{(a)}$, $k^{(b)}$, $k^{(c)}$, and $\varepsilon \ll\Delta k$.
It is important to note that $\Delta k$ has to be smaller than the
characteristic scale of variation of $R(\varepsilon,k)$ as a function of $k$.
We will see later that this scale is ${} \sim
\varepsilon{\bar d}'(k)$, where the prime means derivative with respect to
$k$.  Therefore, the choice of the averaging interval length is limited by
the inequality $ \varepsilon{\bar d}'(k)\Delta k\lesssim 1$.

The fourth ellipse in the diagram Fig.~\ref{fig:dfn.pol} is drawn in the area
where $\varepsilon \sim \Delta k$.  It is difficult to decide whether the
value of $k$ in this case is $k^{(a)}$, or $k^{(b)}$, or maybe $k^{(c)}$. In 
order to make the correlation function well defined for all $\varepsilon$ we
multiplied it by Gaussian prefactor $e^{- \Delta l^2\varepsilon^2/2}$, see
Eq.~(\ref{eq:dfn.3}) where $\Delta l\Delta k\gtrsim1$.

The domain of $\varepsilon$, where the correlation function is meaningful, is
limited now by two competing conditions: $\varepsilon\Delta l\lesssim1$ and
$\varepsilon \lesssim 1/(\Delta k{\bar d}')$. The latter inequality has to be
implied by the former one, and therefore we obtain the important condition
\begin{equation}
   { \Delta k \over \Delta l } \lesssim { 1\over {\bar d}'(k) }\;.
\label{eq:dfn.3a}
\end{equation}
The averaging has to be performed over a large number of the energy levels.
This number is
\begin{equation}
   \Delta {\cal N} \sim \Delta k \bar d(k) \gg 1\;.
\label{eq:dfn.3b}
\end{equation}
The inequalities Eqs.~(\ref{eq:dfn.3a}) and  (\ref{eq:dfn.3b}) are not very
restrictive and most of the published experimental and numerical work, where
the correlation function was computed, employed the averaging intervals of
the width satisfying them.

\begin{figure}
\unitlength=1mm
\begin{picture}(88,72)
\put(00.,00.){\epsfxsize=100mm\epsffile{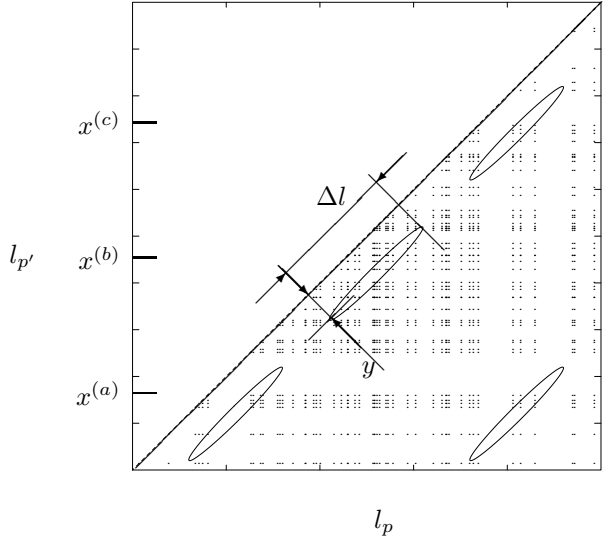}}
\put(22.00,10.00){\line(1,1){62.00}}
\put(41.00,37.00){\vector(1,-1){4.00}}
\put(52.00,26.00){\vector(-1,1){4.00}}
\put(41.00,37.00){\line(1,-1){14.00}}
\put(53.00,23.00){\makebox(0,0)[cc]{$y$}}
\put(45.00,27.00){\line(1,1){6.00}}
\put(38.00,32.00){\line(1,1){17.00}}
\put(39.00,33.00){\vector(1,1){3.00}}
\put(58.00,52.00){\vector(-1,-1){4.00}}
\put(53.00,49.00){\line(1,-1){10.00}}
\put(50.00,45.00){\makebox(0,0)[rb]{$\Delta l$}}
\put(55.00,2.00){\makebox(0,0)[cb]{$l_p$}}
\put(7.00,38.00){\makebox(0,0)[cc]{$l_{p'}$}}
\put(21.70,38.00){\line(1,0){3.00}}
\put(20.00,38.00){\makebox(0,0)[rc]{$x^{(b)}$}}
\put(21.70,20.00){\line(1,0){3.00}}
\put(20.00,20.00){\makebox(0,0)[rc]{$x^{(a)}$}}
\put(21.70,56.00){\line(1,0){3.00}}
\put(20.00,56.00){\makebox(0,0)[rc]{$x^{(c)}$}}
\end{picture}
\caption{
Schematic diagram demonstrating how to compute the correlation
function
of the length spectrum. Each point represents the pair of the
primitive periodic orbits
lengths $(l_{p},l_{p'})$.  The correlation function $\tilde R(x, y)$
picks up the points inside one of the ellipses shown in the diagram. The three
ellipses near the diagonal were drawn for $x$ equal to $x^{(a)}$, $x^{(b)}$,
$x^{(c)}$, and $y \ll \Delta l$. The fourth ellipse was drawn for the case
$y \sim \Delta l$ and the choice of relevant $x$ becomes ambiguous.  }
\label{fig:dfn.pll}\end{figure}

Similar statistics can be defined for the length of the PPOs entering the
right hand side of Eq.~(\ref{eq:dfn.1b}). The length spectrum of the system
has the density $\sum_p \delta(x-l_p)$, and the weighted mean density $
\tilde d_{\text{ppo}}(x) $ is the sum over $n,\nu,\mu$ of
\begin{eqnarray}
   \bar d_{\text{ppo}}(x,n,\nu,\mu)
   &=& 
   \sum_p e^{-\lambda_p l_p} \delta_{n\,n_p}
   \delta_{\nu\,\nu_p} \delta_{\mu\,\mu_p}
   { e^{- {(x-l_p)^2\over 2\Delta l^2}} \over\sqrt{2\pi} \Delta l } 
\nonumber\\
   &\equiv& \bar d_{\text{ppo}}(X) \;,\;\;X=(x,n,\nu,\mu)\;,
\label{eq:dfn.4}
\end{eqnarray}
where  we have set the averaging interval to be precisely $\Delta l$, 
because we are going to build analytical relations between the statistics
of  energy levels and the statistics of  periodic orbits. The mean density in
Eq.~(\ref{eq:dfn.4}) picks up the orbits $p$ with the defined value of the 
number of the wall reflections $n_p$. We keep this number in
Eq.~(\ref{eq:dfn.4}), because the action correlations were found between the
orbits with the same $n_p$, see Ref.\onlinecite{Cohen-Primack-Smilansky}.

\begin{mathletters}
The correlation function of the length spectrum is defined in a slightly
different way from the correlation function of the density of states:
\begin{eqnarray}
   \lefteqn{ \tilde R_2(x,n,\nu,\mu; x',n,\nu',\mu') 
   =\sum_{p} e^{-{\lambda_p l_p\over2}}
  \delta_{n\,n_p}\delta_{\nu\,\nu_p} \delta_{\mu\,\mu_p} 
   }
\nonumber\\
   &\times&
  \biggl\{
   \sum_{p'}
       e^{-{\lambda_{p'} l_{p'}\over 2}}
      \delta_{n\,n_p}\delta_{\nu'\,\nu_{p'}} \delta_{\mu'\,\mu_{p'}}
      { e^{- {(x-l_p)^2\over 2\Delta l^2}}\over\sqrt{2\pi}\Delta l}
      { e^{- {(x'-l_{p'})^2\over 2\Delta l^2}}\over\sqrt{2\pi}\Delta l}
\nonumber\\
   & - &
   \sum_{p'\ne p}
       e^{-{\lambda_{p'} l_{p'}\over2}}
      \delta_{n'\,n_{p'}}\delta_{\nu'\,\nu_{p'}} \delta_{\mu'\,\mu_{p'}}
      \delta(x-x'-l_p+l_{p'})
\nonumber\\
      &\times &
      { e^{- {1\over 2\Delta l^2}({x+x'\over 2}-{l_p+l_{p'} \over 2})^2}
      \over
      \sqrt{2\pi}\Delta l}
   \biggr\}
   e^{-{ \Delta k^2 (x-x')^2/2}}
\nonumber\\
      &\equiv & \tilde R_2(X;X')\;\;,
\label{eq:dfn.5a}
\end{eqnarray}
where the first term in the braces is the product of the mean densities of
PPOs  and the second term is the probability of finding two PPOs having the
length difference $y$. Both terms are weighted by stability factors
which compensate  the exponential proliferation of the PPOs.

The key assumption of the present theory is that the orbits with different
Maslov indexes or different winding numbers or different number of scatterings
do not contribute to the
correlation function of the length spectrum, which we should define as
\begin{equation}
  \tilde R(x,y) = \sum_{n,\nu,\mu}
  \tilde R_2(x+{y\over 2},n,\nu,\mu; x-{y\over 2},n,\nu,\mu)\;. 
\label{eq:dfn.5d}
\end{equation}
Therefore we assume that  $\tilde R_2(X;X')$ decay very fast with $|n-n'|$,
$|\nu-\nu'|$, and $|\mu-\mu'|$. The correlation function $\tilde R_2(X;X')$
is meaningful for $|x-x'|\lesssim \Delta l$ and this is taken into account by
the factor $e^{-{ \Delta k^2 (x-x')^2/2}}$ in Eq.~(\ref{eq:dfn.5a}), since
$1/\Delta k\lesssim \Delta l$ as we will show later. 
\label{eq:dfn.5}
\end{mathletters}

\begin{mathletters}
The correlation function defined by Eq.~(\ref{eq:dfn.5d}) has a
normalization condition which is an extremely useful tool for verifying
results. We have from Eqs.~(\ref{eq:dfn.5})
\begin{eqnarray}
   \sum_{n'\,\nu'\,\mu'}  \int dx' \,
   \tilde R_2(X;X')  &=&   \bar d_{\text{ppo}}(X)
\\
    \int dy \, \tilde R(x,y)  &=&   \tilde d_{\text{ppo}}(x)
\end{eqnarray}
which are valid under the condition $\Delta l\ll x$. In order to derive this
relation we have replaced $\sum_{p}\sum_{p'\ne p}$ in Eq.~(\ref{eq:dfn.5a}) by 
two terms $\sum_{pp'} - \sum_p\delta_{pp'}$. The second term gives the mean
density on the right hand side of Eqs.~(\ref{eq:dfn.5b}). The first term is a
product of the mean densities and cancels another term of the type $\sum_{pp'}$
in Eq.~(\ref{eq:dfn.5a}).    
\label{eq:dfn.5b}
\end{mathletters}

The correlation functions defined by Eqs.~(\ref{eq:dfn.5})  have the
graphical representation in Fig.~\ref{fig:dfn.pll} quite similar to that in
Fig.~\ref{fig:dfn.pol}.  In the case of Fig.~\ref{fig:dfn.pll} we are
interested in the number of PPO pairs whose mean length is near $x$ and 
difference of lengths is near $y$. Correlation function counts the number of
PPO pairs inside one of the ellipses shown in Fig.~\ref{fig:dfn.pll}. The
``length'' of the ellipses is determined by the width of the Gaussians in
Eq.~(\ref{eq:dfn.5a}) and to be $ \sim \Delta l$. The ``width'' of the
ellipses does not appear explicitly in Eq.~(\ref{eq:dfn.5a}), meaning that
$\delta(y-l_p+l_{p'})$ has to be smoothed to  the order of  $1/\tilde
d_{\text{ppo}}(x)$.

The length of the averaging interval, $\Delta l$, has to be as large as
possible, but smaller than the characteristic scale of the variation of
$\tilde R(x,y)$ as a function of $x$ for constant $y$. This scale can be
estimated from the results of Ref.\onlinecite{Argaman-dec93} and it is ${\bar
d}'(k)/y$ taken for $k=k_x$, where $k_x$ is the inverse density of states:
\begin{equation}
   x = 2\pi \bar d(k_x)\;,
\label{eq:disc.7}
\end{equation}
which is a classical quantity. The choice of the averaging interval  is,
therefore, limited by the inequality $\Delta l \lesssim {\bar d}'(k)/y$. This
inequality has to be fulfilled in the interval of $y$, where the correlation
function Eq.~(\ref{eq:dfn.5}) is meaningful, i.e. for $0<y\lesssim 1/\Delta
k$. Therefore, the
choice of the averaging window width is limited by the condition analogous to
Eq.~(\ref{eq:dfn.3b})
\begin{equation}
   {\Delta l \over \Delta k}
   \lesssim  {\bar d}'(k_x)\;.
\label{eq:dfn.5c}
\end{equation}
This inequality holds regardless of Eq.~(\ref{eq:dfn.3b}) because one
can study the statistics of energy levels independently from the statistics
of periodic orbits. From this point of view the inequality
Eq.~(\ref{eq:dfn.5c}) is not very restrictive either.

\begin{mathletters}
The statistical properties of the PPOs length spectrum are ultimately connected
with the statistical properties of the quantum mechanical
spectrum\cite{Argaman-dec93}. We obtain from Eqs.~(\ref{eq:dfn.2}) the
periodic orbit expansion of the spectral correlations form-factor
\begin{eqnarray}
   K(x,k) &\equiv&
   \int_{-\infty}^\infty
   { d\varepsilon \over 2 \pi }
   e^{-i\varepsilon x} R(\varepsilon,k)
   =  K_{\text{diag}}(x)+ K_{\text{off}}(x,k)
\nonumber\\
\relax
\label{eq:dfn.6a}
\\
   K_{\text{diag}}(x)
   &=&
   { 1 \over 4\pi^2 }
   \sum_{pr}
   { l_p^2 
   \over |e^{\lambda_p r l_p \over 2}+e^{-{\lambda_p r l_p \over 2}}|^2}
   {
      e^{- {1\over 2\Delta l^2} (|x|- l_pr)^2 }
         \over
      \sqrt{2\pi} \Delta l
   }
\nonumber\\
\relax
\label{eq:dfn.6b}
\\
   K_{\text{off}}(x,k)
   &=&
  \sum_{p}\sum_{p'\ne p}
   {
      l_p l_{p'}
      e^{- {\lambda_pl_p+\lambda_{p'}l_{p'}\over 2}}
      \over
      4\pi^2 }
   {
      e^{- {1\over 2\Delta l^2} (|x|-{l_p+l_{p'}\over 2})^2 }
   \over \sqrt{2\pi } \Delta l}
\nonumber\\
   &\times&
   e^{ik(l_p - l_{p'}) - \Delta k^2(l_p-l_{p'})^2 /2}
\nonumber\\
   &\times&
   e^{ i(\nu_p-\nu_{p'})\pi/2 + i\phi(\mu_p - \mu_{p'})}
\label{eq:dfn.6d}
\\
   &=&
   -{1\over 4\pi^2 }\, x^2 
   \int_{-\infty}^\infty
      dy\,e^{iky}
      \tilde R(|x|,y)
\label{eq:dfn.6e}
\end{eqnarray}
which is valid for $k\gg \Delta k$. Under this condition we have neglected 
the Fourier transform of the first term in the braces on the left hand side of
Eq.~(\ref{eq:dfn.5a}), because it contains the factor $e^{-k^2/(2\Delta k^2)}$.
\label{eq:dfn.6}
\end{mathletters}

Equation~(\ref{eq:dfn.6d}) is justified for $x$ such that
\begin{equation}
   \lambda |x| \gg 1\;,
\label{eq:dfn.6f}
\end{equation}
where $\lambda$ is the mean Lyapunov exponent defined per unit length of the
trajectory.
This condition allows us to expand the denominators in Eq.~(\ref{eq:dfn.1b})
and to neglect the summation over the repetition index $r$. We also assumed
that the magnetic flux $\phi$ is so large that $\phi\Delta \mu \gg1$, where
$\Delta\mu$ is the characteristic scale of $\mu-\mu'$-dependence of 
$\tilde R_2(X;X') $.

The key question of the theory is how to perform the summation over the
periodic orbits if the length spectrum is unknown. This question  was
resolved\cite{Berry-feb85} with the help of the sum
rule\cite{Hannay-OzorioDeAlmeida}, valid for ergodic systems.  One can deduce
the density of PPOs  from this sum rule:
\begin{equation}
   \tilde d_{\text{ppo}}(x) \sim { 1 \over x}\;.
\label{eq:dfn.7}
\end{equation}
This expression is valid asymptotically for $x$ larger than the ergodic length
($\tau_{\text{erg}}$ times velocity) and it gives $K_{\text{diag}}(x)\propto x$.

The definition of the form-factor, Eqs.~(\ref{eq:dfn.6}), is not complete,
because we did not specify yet how to choose the sizes of the averaging
windows $\Delta k$ and $\Delta l$. Common wisdom was that
Eqs.~(\ref{eq:dfn.6}) would survive if they do not break the conditions
\begin{equation}
   \Delta l e^{\lambda x} / x \gg1 \;,\;\;\;\;
   \Delta k \bar d(k) \gg 1\;.
\label{eq:dfn.8}
\end{equation}
We want to define the form-factor in such a way that it will carry
information about the correlations of energy levels and about the correlations
of periodic orbits. Therefore, we require $\Delta k$ and $\Delta l$ to
satisfy Eqs.~(\ref{eq:dfn.3b}) and (\ref{eq:dfn.5c}) i.e.
\begin{equation}
     \bar d'(k)
     \lesssim {\Delta l \over \Delta k} \lesssim
     \bar d'(k_x) \;.
\label{eq:dfn.9}
\end{equation}
Near the Heisenberg length $x\sim 2\pi \bar d(k)$, we have $k_x = k$ and two
inequalities in Eq.~(\ref{eq:dfn.9}) give us $\Delta l / \Delta k \sim \bar
d'(k)$.


\section{Definitions of objects related to the spectral zeta function.}
\label{sec:defZfun}

One can construct an infinitely large number of complex analytic functions of
$k$ having zeros only at $k=k_n$. Each such function can be considered as a
spectral determinant. We are going to define the spectral determinant in 
terms of the retarded Green function $\hat G^r(k)$. The trace of this
operator is an  analytic function and we have
\begin{equation}
   \zeta(k) \equiv
   e^{ \int_0^k dq
   \left\{
      \text{Tr} \hat G^r(q)E'(q) + i\pi\bar d(q)
   \right\}}\;.
\label{eq:dfz.1}
\end{equation}
This definition implies the functional equation for the spectral zeta function
\begin{equation}
   \zeta(k) = e^{ 2\pi i\int_0^k dq\bar d(q) }\zeta^\ast(k)\;,
\label{eq:dfz.2}
\end{equation}
where the asterisk stands for the complex conjugation. The expression in 
braces in Eq.~(\ref{eq:dfz.1}) can be written as the series expansion over 
the periodic orbits Eq.~(\ref{eq:dfn.1b}). The exponential function of this
series was computed by Voros\cite{Voros-88} and the result is
\begin{equation}
  \zeta(k) =
  \prod_{p}\prod_{r=0}^\infty
  \left( 1 - e^{i(kl_p -\nu_p\pi/2+\phi\mu_p)}
   e^{-(r+1/2) \lambda_p l_p} \right)\;.
\label{eq:dfz.3}
\end{equation}
This expression is taken from Eq.~(6.21) of Ref.\onlinecite{Smilansky-rev95},
and $\lambda_p$ here is the Lyapunov exponent per unit length. Here we added 
the phase due to the Aharonov -- Bohm flux.

The autocorrelation function of the spectral zeta is defined similarly to the
correlation function of the density of states:
\begin{eqnarray}
  C(\varepsilon,k)
  &\equiv&
  e^{-\Delta l^2\varepsilon^2/2}
  \,
  \int_{-\infty}^\infty
  {
     dq  \over
     \sqrt{ 2 \pi }\Delta k
  }
  \,
  e^{- {(q-k)^2 \over 2\Delta k^2 }}
\nonumber\\
  &\times&
  \zeta(q+{\varepsilon\over 2})
  \zeta^\ast(q-{\varepsilon\over 2})
\label{eq:dfz.4}
\end{eqnarray}
Equation.~(\ref{eq:dfz.2}) is also valid for the correlation
function
\begin{equation}
   C(\varepsilon,k) = e^{ 2\pi i \bar d(k) \varepsilon } 
   C(-\varepsilon,k)\;,
\label{eq:dfz.4a}
\end{equation}
where we expanded the integral in the exponent in Eq.~(\ref{eq:dfz.2})
under the condition $\varepsilon\ll k$, and also used the inequality
Eq.~(\ref{eq:dfn.3b}).


As usual, we put the prefactor in Eq.~(\ref{eq:dfz.4}), which will be
translated into the averaging over the lengths of the CPOs. The CPOs appear
from the evaluation of the product in Eq.~(\ref{eq:dfz.3}) and we define
them as all possible sets of the PPOs taken without repetitions:
\begin{equation}
   c = \{p\}\text{   and    } l_c\equiv \sum_{p\in c} l_p\;.
\label{eq:dfz.5f}
\end{equation}
The CPOs have the composite Lyapunov exponents
$
   \lambda_c \equiv {1\over l_c} \sum_{p\in c} \lambda_p l_p
$,
the composite Maslov indexes
$
   \nu_c \equiv \sum_{p\in c} \nu_p
$,
the composite number of wall reflections
$
   n_c \equiv \sum_{p\in c} n_p
$,
the number of the primitive components
$
   m_c \equiv \sum_{p\in c} 1
$,
and the composite winding numbers
$
   \mu_{c}\equiv \sum_{p\in c} \mu_p
$.
This definition of the CPOs leads to the high degeneracy of the
composite length spectrum $\{l_c\}$.

The composite length spectrum has the density $\sum_c\delta(x-l_c)$ and
the weighted mean density $\tilde d_{\text{cpo}}(x) $ is the sum over 
$n,\nu,\mu$ of
\begin{eqnarray}
   \bar d_{\text{cpo}}(x,n,\nu,\mu)
   &=&   \sum_c e^{-\lambda_c l_c}
   \delta_{n\,n_p} \delta_{\nu\,\nu_p} \delta_{\mu\,\mu_p}
   { e^{- {(x-l_c)^2\over 2\Delta l^2}}\over\sqrt{2\pi}\Delta l} 
\nonumber\\
   &\equiv& \bar d_{\text{cpo}}(X) \;.
\label{eq:dfz.6}
\end{eqnarray}
The correlation function of the composite length spectrum is defined similarly
to Eq.~(\ref{eq:dfn.5})
\begin{eqnarray}
   \lefteqn{ \tilde C_2(x,n,\nu,\mu; x',n,\nu',\mu') 
   =\sum_{c} e^{-{\lambda_c l_c\over2}}
  \delta_{n\,n_c}\delta_{\nu\,\nu_c} \delta_{\mu\,\mu_c} 
   }
\nonumber\\
   &\times&
  \biggl\{
   \sum_{c'}
       e^{-{\lambda_{c'} l_{c'}\over 2}}
      \delta_{n\,n_c}\delta_{\nu'\,\nu_{c'}} \delta_{\mu'\,\mu_{c'}}
      { e^{- {(x-l_c)^2\over 2\Delta l^2}}\over\sqrt{2\pi}\Delta l}
      { e^{- {(x'-l_{c'})^2\over 2\Delta l^2}}\over\sqrt{2\pi}\Delta l}
\nonumber\\
   & - &
   \sum_{c'\ne c}
       e^{-{\lambda_{c'} l_{c'}\over2}}
      \delta_{n'\,n_{c'}}\delta_{\nu'\,\nu_{c'}} \delta_{\mu'\,\mu_{c'}}
      \delta(x-x'-l_c+l_{c'})
\nonumber\\
      &\times &
      { e^{- {1\over 2\Delta l^2}({x+x'\over 2}-{l_c+l_{c'} \over 2})^2}
      \over
      \sqrt{2\pi}\Delta l}
   \biggr\}
   e^{-{ \Delta k^2 (x-x')^2/2}}
\nonumber\\
      &\equiv & \tilde C_2(X;X')\;\;,
\label{eq:dfz.7}
\\  \lefteqn{
    \tilde C(x,y) = \sum_{n,\nu,\mu}\,
    \tilde C_2(x+{y\over 2},n,\nu,\mu; x - {y\over 2},n,\nu,\mu)\;.
    }
\label{eq:dfz.7b}
\end{eqnarray}

\begin{mathletters}
We will assume in what follows that $\tilde C_2(X;X')$ falls off rapidly with
$|n-n'|$, $|\nu-\nu'|$ and $|\mu-\mu'|$. The correlation function $\tilde
C(x,y)$ is meaningful only for $y\lesssim \Delta l$, and this is taken into
account by the Gaussian prefactor. The normalization ofthe  correlation
function can be deduced from Eqs.~(\ref{eq:dfz.6}) and (\ref{eq:dfz.7})
\begin{eqnarray}
   \sum_{n'\,\nu'\,\mu'}\int_{0}^\infty dx'\, \tilde C_2(X;X')
   &=&
   \bar d_{\text{cpo}}(X)
   \;,
\\
   \int_{-\infty}^\infty dy\, \tilde C(x,y)
   &=&
   \tilde d_{\text{cpo}}(x)
   \;,
\end{eqnarray}
and it is valid under the condition $\Delta l\ll x$. The derivation is similar
to the derivation of Eq.~(\ref{eq:dfn.5b}).
\label{eq:dfz.7a}
\end{mathletters}

\begin{mathletters}
The statistical properties of the CPOs length spectrum are connected with the
statistical properties of the spectral zeta function. We can see that by
expanding the Fourier transform of the correlation function
Eq.~(\ref{eq:dfz.4}) over the CPOs
\begin{eqnarray}
   K^\zeta(x,k) &\equiv&
   \int_{-\infty}^\infty
   { d\varepsilon \over 2 \pi }
   e^{-i\varepsilon x} C(\varepsilon,k)
   =  K^\zeta_{\text{diag}}(x)+ K^\zeta_{\text{off}}(x,k)
\nonumber\\
\relax
\label{eq:dfz.8a}
\\
   K^\zeta_{\text{diag}}(x)
   &=&
   \sum_{c} \sum_{r_p,\,p\in c}
   \left\{\prod_{p\in c}
   {   e^{-\lambda_pl_pr_p^2}
       \over 
       \prod_{j=1}^{r_{p}}(1-e^{-\lambda_pl_pj})^2
       }\right\}
\nonumber\\
   &\times&
   {  1
         \over
      \sqrt{2\pi} \Delta l^2
   }
      e^{-  { \left(x- \sum_{p\in c} l_p r_p \right)^2 / ( 2\Delta l^2)} }
\label{eq:dfz.8bb}
\\
   &\approx& 
   \tilde d_{\text{cpo}}(x)
\label{eq:dfz.8b}
\\
   K^\zeta_{\text{off}}(x,k)
   &\approx&
  \sum_{c}\sum_{c'\ne c}
      e^{- {\lambda_cl_c+\lambda_{c'}l_{c'}\over 2}}
   {
      e^{- {1\over 2\Delta l^2} (|x|-{l_c+l_{c'}\over 2})^2 }
   \over \sqrt{2\pi}\Delta l^2}
\nonumber\\
   &\times&
   e^{ik(l_c- l_{c'})- \Delta k^2(l_c-l_{c'})^2/2 }
   e^{i\pi(m_c-m_c')}
\nonumber\\
   &\times&
   e^{ i(\nu_c-\nu_{c'})\pi/2 }
   e^{i\phi(\mu_{c} - \mu_{c'})}
\label{eq:dfz.8d}
\\
   &=&
   -
   \int_{-\infty}^\infty
      dy\,e^{iky}
      \tilde C(x,y)\;,
\label{eq:dfz.8c}
\end{eqnarray}
where the repetition indexes $r_p$ in Eq.~(\ref{eq:dfz.8bb}) run from $1$ to
$\infty$. Equations~(\ref{eq:dfz.8b}), (\ref{eq:dfz.8d}) and
(\ref{eq:dfz.8c}) are valid under the condition Eq.~(\ref{eq:dfn.6f}).
This condition allows us to  keep only the factors with $r=0$ in
Eq.~(\ref{eq:dfz.3}), and therefore our definition of the CPOs
Eq.~(\ref{eq:dfz.5f}) does not contain the repetitions of the PPOs. We also
assumed that the magnetic flux  $\phi$ is so large that $\phi\Delta \mu
\gg1$, where $\Delta\mu$ is the characteristic scale of  the
$\mu-\mu'$-dependence of $\tilde C_2(X;X')$. Equation~(\ref{eq:dfz.8a})
defines the form-factor of the autocorrelation function of the spectral
zeta.  It is zero for negative $x$, whereas $K(x)$ depends on $|x|$.
\end{mathletters}

The averaged density of the CPOs length spectrum was
computed\cite{Aurich-Steiner-92} for a family of strongly chaotic systems and
its asymptotic form in the general case is
\begin{eqnarray}
  \tilde d_{\text{cpo}}(x)
  \sim  \gamma \;,
\label{eq:dfz.9}
\end{eqnarray}
where $\gamma$ is the normalization constant, and $\gamma^{-1}$ is of the
order of the billiard size. This density is sufficient for the computation of
the diagonal part of the form-factor Eq.~(\ref{eq:dfz.8b}) but the
non-diagonal part can be computed only if the correlations in the CPOs are
known.


\section{Probabilistic theory of the length spectrum correlations.}
\label{sec:pnt}

The length spectrum of PPOs is a part of the length spectrum of the CPOs.
Therefore we can ask what is the probability to find a PPO in some interval
of the composite length spectrum. Let $\bf P$ be the set of all PPOs and let
$\bf C$ be the set of all CPOs.  Then ${\bf P} \subset {\bf C} $. It may
happen that $c\in {\bf C}$ consists of only one PPO and therefore $c\in
{\bf P}$. We define the probability of this event
\begin{equation}
   P(c)
      \equiv
   \Bigl\langle  c\in {\bf P} \Bigr\rangle_{l_c}
       =
     {\bar d_{\text{ppo}}(X_c) \over \bar d_{\text{cpo}}(X_c)}
     \;.
\label{eq:pnt.1}
\end{equation}
It is a function of $X_c$ where $X_c=(l_c,n_c,\nu_c,\mu_c)$. The probability
$P(c)$ counts CPOs, which are actually PPOs, having defined values of $n_c$,
$\nu_c$, $\mu_c$ and lengths which lie in the small interval near $l_c$. In
Eq.~(\ref{eq:pnt.1}) we regard the statement $c\in {\bf P}$ as a boolean
function, which is equal to one if $c\in {\bf P}$ and equal to zero if
$c\notin {\bf P}$.

We can also count the number of pairs of the  CPOs,  $c,c'\in{\bf C}$ with
the fixed length difference $l_c-l_{c'}$, such that both of them are actually
PPOs, $c,c'\in{\bf P}$. The probability of finding such a pair is defined as
\begin{eqnarray}
   P(c,c')
      & \equiv  &
   \Bigl\langle
      c\in {\bf P} \;\cap\; c'\in {\bf P}
   \Bigr\rangle_{l_c+l_{c'}\over 2}
   = P(c)P(c')
\nonumber\\
   &-& 
   e^{-{\lambda_c l_c + \lambda_{c'}l_{c'}\over 2}}
   \,{\tilde R_2(X_c; X_{c'})
      \over
      \bar d_{\text{cpo}}(X_c) \bar d_{\text{cpo}}(X_{c'})
   }\;.
\label{eq:pnt.2}
\end{eqnarray}
This equation defines the averaging over $l_c+l_{c'}\over 2$. The fluctuations
of the Lyapunov exponent are ignored here for two reasons: First, it was found
numerically\cite{Cohen-Primack-Smilansky} that the fluctuations of the
stability amplitudes do not affect the correlation function; second, our
result Eq.~(\ref{eq:pnt.12}) is independent of these fluctuations, because
$P(c)$ is smooth function of $X_c$.

The probability that CPO $c$ contains another CPO $c'$ can be defined
by averaging over the length of $c$
\begin{equation}
   P(c'\in c)
      \equiv
   \Bigl\langle c'\in c \Bigr\rangle_{l_c}
   = e^{-\lambda_{c'}l_{c'}}\,
   { \bar d_{\text{cpo}}(X_c-X_{c'}) \over \bar d_{\text{cpo}}(X_c) }
   \;.
\label{eq:pnt.3}
\end{equation}
This probability is related to the probability $P(c)$ in Eq.~(\ref{eq:pnt.1})
since
\begin{equation}
   c \in {\bf P}
   \;\; \Leftrightarrow  \;\;
   \bigcap_{p\in{\bf P},\;l_p<l_c^\ast}\;\;
   p\notin c\;,
\label{eq:pnt.4}
\end{equation}
where $l_c^\ast$ is an arbitrary length such that
$l_c/2 \ge l_c^\ast < l_c$. Averaging over $l_c$ gives
\begin{eqnarray}
   P(c) &=& \prod_{p,\; l_p < l_{c}^\ast}
   \Bigl[ 1 - P(p\in c) \Bigr]
\nonumber\\
    &+&
    \sum_{c'}
    (-1)^{m_{c'}}
    \Bigl[ P(c'\in c) - \prod_{p\in c'}P(p\in c) \Bigr]
    \;,
\label{eq:pnt.5}
\end{eqnarray}
where the composite orbits $c'$ are such that $l_{c'}\ge l_c^\ast$ and
$\forall p\in c'\;\;l_p < l_c^\ast$. We assume here that $P(c'\in
c)=\prod_{p\in c'} P(p\in c)$ unless $l_{c'}$ is larger than $l_{c}^\ast$.
One can choose $l_c^\ast$ by minimizing the absolute value of the second term
on the right hand side of Eq.~(\ref{eq:pnt.5}).

The main purpose of this section is to construct an equation analogous to
Eq.~(\ref{eq:pnt.5}) for the joint probability $P(c,c')$.
Equation~(\ref{eq:pnt.4}) for pair of CPOs reads
\begin{equation}
   c\in P \;\cap\;c'\in P
   \;\; \Leftrightarrow  \;\;
   \bigcap_{p\in{\bf P},\;\;l_p<l_c^\ast}\;\;
   p\notin c \;\cap\;p\notin c'\;.
\label{eq:pnt.6}
\end{equation}
This statement can be converted to a Boolean expression and averaged over
$(l_c+l_{c'}) / 2$. That is
\begin{eqnarray}
   P(c,c') =&
   \biggl\langle
      \prod_{p,\, l_p < l_{c}^\ast}
   \Bigl[
   1 &- \;p\in c\; - \;p\in c'\;
\nonumber\\
   &&+ \;p\in c\;\cap\;p\in c'\;\Bigr]
   \biggr\rangle_{l_c+l_{c'}\over 2}
\label{eq:pnt.7a}
\end{eqnarray}
Let CPO $c''$ consist only of the PPOs which are shorter than $l_c^\ast$.
We can introduce the sum over such orbits into Eq.~(\ref{eq:pnt.7a})
\begin{eqnarray}
   \lefteqn{
   P(c,c')
   =
   \biggl\langle
      \sum_{c''}
         \prod_{p\notin c'',\, l_p < l_{c}^\ast}
   \Bigl[
      1 - \;p\in c\; - \;p\in c'
   \Bigr]
   }
\nonumber\\
   &\times&
         \prod_{p\in c''}
   \Bigl[p\in c\;\cap\;p\in c'\;\Bigr]
   +
   \prod_{l_p < l_{c}^\ast}
      \Bigl[  1 - \;p\in c\; - \;p\in c' \Bigr]
   \biggr\rangle
\nonumber\\
\relax\label{eq:pnt.7b}
\\
   &\approx&
      \sum_{c''}
      \prod_{p\notin c'',\, l_p < l_{c}^\ast}
      \Bigl[  1 - P(p\in c) - P(p\in c') \Bigr]
\nonumber\\
   &\times&
         P(c''\in c\;\cap\;c''\in c')
   + \prod_{l_p < l_{c}^\ast}
      \Bigl[  1 - P(p\in c) - P(p\in c') \Bigr]\;,
\nonumber\\
\relax\label{eq:pnt.7}
\end{eqnarray}
where the probability of having a ``common divisor'' is given by
\begin{equation}
   P(c''\in c\;\cap\;c''\in c')
      \equiv
   \Bigl\langle c''\in c\;\cap\;c''\in c'\Bigr\rangle_{l_c+l_{c'}\over 2}
    \;.
\label{eq:pnt.9}
\end{equation}
We assumed  that for $p\ne p'$
\begin{equation}
  \Bigl\langle
     p\in c\;\cap\;p'\in c'
  \Bigr\rangle_{l_c+l_{c'}\over 2}
  \approx
  \Bigl\langle
     p\in c
  \Bigr\rangle_{l_c}
  \Bigl\langle
     p'\in c'
  \Bigr\rangle_{l_{c'}}
\label{eq:pnt.10}
\end{equation}
and also we made the same assumptions as in the derivation of
Eq.~(\ref{eq:pnt.5}). Particularly we neglected correction terms in
Eq.~(\ref{eq:pnt.7}), which are sums over $c'',\, l_{c''}\ge l_c^\ast$.
Such correction terms were important in Eq.~(\ref{eq:pnt.5}), but they can
be neglected in Eq.~(\ref{eq:pnt.7}) because the greatest contribution to the
correlation of PPOs is given by the short CPOs $c''$.

It is convenient to rewrite  Eq.~(\ref{eq:pnt.7}) as a relation between the
correlation function of PPOs and CPOs having a ``common divisor''
\begin{eqnarray}
   \lefteqn{
      P(c)P(c')-P(c,c')
   }
\nonumber\\
   &\approx&
   \sum_{c''}\biggl\{\Bigl[
   P(c''\in c)P(c''\in c')
   -
   P(c''\in c\,\cap\,c''\in c')
   \Bigr]
\nonumber\\
   &&\times
   \prod_{p\notin c'',\, l_p < l_{c}^\ast}
   \Bigl[
      1 - P(p\in c) - P(p\in c')
   \Bigr]\biggr\}\;.
\label{eq:pnt.8}
\end{eqnarray}
It is essential for the derivation of Eq.~(\ref{eq:pnt.8}) that the composite
orbit $c''$ does not contain repetitions of the primitive orbits. This agrees
with our definition of composite orbits Eq.~(\ref{eq:dfz.5f}). In general
one can allow repetition of the primitive orbits in the definition of the
composite orbit. In this case Eq.~(\ref{eq:pnt.8}) is also correct, but the
sum must be taken over the composite orbits which does not contain repetitions
of the primitive orbits.

The derivation of Eq.~(\ref{eq:pnt.8}) contains a number of approximations.
Therefore, one would like to verify that this equation preserves the
normalization of the correlation functions. This verification appears in 
Appendix~\ref{sec:Norm}. 

In certain cases, any given composite orbit contains a small number of 
PPOs and therefore we
can make a further approximation in Eq.~(\ref{eq:pnt.8})
\begin{eqnarray}
   &&\prod_{p\notin c'',\, l_p < l_c^\ast}
   \Bigl[
      1 - P(p\in c) - P(p\in c')
   \Bigr]
\nonumber\\
    &\approx&
   \prod_{l_p < l_c^\ast}
   \Bigl[ 1 - P(p\in c) \Bigr]
   \Bigl[ 1 - P(p\in c') \Bigr]\;.
\label{eq:pnt.10a}
\end{eqnarray}
The condition of a small number of ``divisors''
\[   P(p\in c) \propto \, e^{-\lambda_p l_p}\;\;\ll\;\; 1      \]
is not enough to justify making this approximation. It might
happen that the error becomes larger when the sum over $c''$ in
Eq.~(\ref{eq:pnt.8}) is computed. One can check that this error is of the same
order as the contribution of the repetitions of PPOs which has been neglected.

From Eqs.~(\ref{eq:pnt.8}) and (\ref{eq:pnt.10a}) by making
use of Eq.~(\ref{eq:pnt.5}) we obtain 
\begin{eqnarray}
   \lefteqn{
      P(c)P(c')-P(c,c')
      \approx
      P(c)P(c')
   }
\nonumber\\
   &\times&
   \sum_{c''}\Bigl[
   P(c''\in c)P(c''\in c')
   -
   P(c''\in c\,\cap\,c''\in c')
   \Bigr]\;.
\nonumber\\
\relax
\label{eq:pnt.11}
\end{eqnarray}
The right hand side of Eq.~(\ref{eq:pnt.8}) or Eq.~(\ref{eq:pnt.11}) contains
the non-trivial probability of finding a ``common divisor'' $P(c''\in
c\,\cap\,c''\in c')$. It is difficult to find a ``common divisor'' if the lengths
of $c$ and $c'$ are close to each other. We can compute this probability by
noting that in order to find a ``common divisor'' of length $l_{c''}$ one has
to find two CPOs $c^{iii}$ and $c^{iv}$ of length 
$l_{c^{iii}}=l_{c}-l_{c''}$ and $l_{c^{iv}}=l_{c'}-l_{c''}$. The same is true
for other components of $X_c$ characterizing each CPO. The probability of
finding a pair of composite orbits of specific lengths is given by the
two-point correlation function of the composite length spectrum. Therefore,
we substitute
\begin{eqnarray}
   \lefteqn{P(c''\in c\;\cap\;c''\in c')
    = P(c''\in c)P(c''\in c')  }
\nonumber\\
   & - &
   e^{-{\lambda_c l_c + \lambda_{c'}l_{c'} \over 2}
   - \lambda_{c''}l_{c''}  }\,
   { \tilde C_2(X_c-X_{c''}, X_{c'}-X_{c''})
   \over \bar d_{\text{cpo}}(X_c) \bar d_{\text{cpo}}(X_{c'})
   }
    \;.
\label{eq:pnt.12a}
\end{eqnarray}
into Eq.~(\ref{eq:pnt.11}) and arrive at
\begin{eqnarray}
   \lefteqn{
      \tilde R_2(X_c, X_{c'}) \approx  P(c)P(c')
   }
\nonumber\\
      & \times &
   \sum_{c''} e^{- \lambda_{c''}l_{c''} }\,
      \tilde C_2(X_c-X_{c''}, X_{c'}-X_{c''})\;.
\label{eq:pnt.12}
\end{eqnarray}
In the more complicated cases, when the approximation Eq.~(\ref{eq:pnt.10a})
fails, we should substitute Eq.~(\ref{eq:pnt.12a}) into Eq.~(\ref{eq:pnt.8})
\begin{eqnarray}
   \lefteqn{
      \tilde R_2(X_c, X_{c'}) \approx  P(c)P(c')
   }
\nonumber\\
      & \times &
     \left\{\prod_{p,\, l_p < l_c^\ast}
     { 1 - P(p\in c) - P(p\in c') \over
      \Bigl[ 1 - P(p\in c) \Bigr]
      \Bigl[ 1 - P(p\in c') \Bigr]
      } \right\}
\nonumber\\
      & \times &
   \sum_{c''}
   {  e^{- \lambda_{c''}l_{c''} }\,
      \tilde C_2(X_c-X_{c''}, X_{c'}-X_{c''})
    \over \prod_{p\in c''}
       \Big[ 1 - P(p\in c) - P(p\in c') \Bigr]
    }  
      \;,
\label{eq:pnt.12b}
\end{eqnarray}
where the product in the braces is assumed to be convergent to some smooth
function of both $X_c$ and $X_c'$ but independent of the $l_c^\ast$.

The substitution of Eq.~(\ref{eq:pnt.1}) into Eq.~(\ref{eq:pnt.12}) gives 
\begin{eqnarray}
   \lefteqn{
      \tilde R_2(X, X')
      =
      { \bar d_{\text{ppo}}(X)\bar d_{\text{ppo}}(X') 
      \over 
      \bar d_{\text{cpo}}(X) \bar d_{\text{cpo}}(X')}
   \int_0^\infty dx'' 
   }
\nonumber\\
   &\times& \sum_{n''\,\nu''\,\mu''}
   \,\bar d_{\text{cpo}}(X'')
   \,\tilde C_2(X-X'', X'-X'') \;,
\label{eq:pnt.14a}
\end{eqnarray}
where the summation in Eq.~(\ref{eq:pnt.12}) is replaced by the integration.

Assuming that the product of the mean densities in Eq.~(\ref{eq:pnt.14a})
varies slowly on the scale of the $X-X'$ dependence of the correlation functions
we can apply the normalization conditions Eqs.~(\ref{eq:dfn.5b}) and
(\ref{eq:dfz.7a}) to get
\begin{eqnarray}
   \lefteqn{
     \bar d_{\text{ppo}}(X)
      =
      \left[{ \bar d_{\text{ppo}}(X) 
      \over 
      \bar d_{\text{cpo}}(X) }
      \right]^2
    }
\nonumber\\
   &\times& \sum_{n'\,\nu'\,\mu'}
   \int_0^\infty dx'\,\bar d_{\text{cpo}}(X')\, 
   \bar d_{\text{cpo}}(X-X')\;,
\label{eq:pnt.15}
\end{eqnarray}
which is satisfied by
\begin{eqnarray}
    \bar d_{\text{ppo}}(X) 
    &\propto&
    {1\over x} 
    { e^{-{\mu^2a\over 2x}} \over \sqrt{2\pi x/a}}
    \,{ e^{-{\nu^2b\over 2x}} \over \sqrt{2\pi x/b}}\;,
\label{eq:pnt.15b}
\\
    \bar d_{\text{cpo}}(X) 
    &\propto&
    \gamma 
    { e^{-{\mu^2a\over 2x}} \over \sqrt{2\pi x/a}}
    \,{ e^{-{\nu^2b\over 2x}} \over \sqrt{2\pi x/b}}\;.
\label{eq:pnt.15c}
\end{eqnarray}
The dependence on $n$ is not known but assumed to be the same for both $
\bar d_{\text{ppo}}(X) $ and $\bar d_{\text{cpo}}(X)$. The parameters $a$ and
$b$ here are lengths of the order of the billiard size. The mean density
of PPOs Eq.~(\ref{eq:pnt.15b}) was introduced by Berry and
Keating\cite{Berry-Keating-may94}. The numeric tests of Dittrich {et
al}\cite{Dittrich-97} support the distribution Eq.~(\ref{eq:pnt.15b}). Then 
the mean density Eq.~(\ref{eq:pnt.15c}) is the solution of
Eq.~(\ref{eq:pnt.15}).

Assuming that $\bar d_{\text{ppo}}(X)/\bar d_{\text{cpo}}(X) = ( \gamma
x)^{-1}$ for large $x$, we obtain the main result of this section
\begin{equation}
      \tilde R(x,y)
      =
   {1 \over x^2\gamma^2}\int_0^\infty dx' \,
   \tilde d_{\text{cpo}}(x')
   \,\tilde C(x-x', y) \;,
\label{eq:pnt.14}
\end{equation}
which was obtained by the summation on  both sides of Eq.~(\ref{eq:pnt.14a})
over $n,\nu,\mu$ at $n'=n$, $\nu'=\nu$, and $\mu'=\mu$.

It is instructive to check how Eqs.~(\ref{eq:pnt.8}) and (\ref{eq:pnt.11})
work for logarithms of integer and prime numbers. It turns out that
Eq.~(\ref{eq:pnt.8}) is precisely equivalent to the Hardy -- Littlewood
conjecture.  This can be shown by simple algebra and we put this calculation
in Appendix~\ref{sec:RZexample}. At the same time Eq.~(\ref{eq:pnt.11}) is
equivalent to the smoothed form of the Hardy -- Littlewood expression, see
Appendix~\ref{sec:RZsmoothed}.


\section{ Behavior of the form-factor near Heisenberg time. }
\label{sec:uff}

For the case of billiards it is convenient to introduce the Heisenberg length
$ 2\pi\bar d(k) $.  From Eqs.~(\ref{eq:dfn.6b}) and (\ref{eq:dfz.8b}) one can
see that the diagonal parts of the form-factors are independent of the
Heisenberg length, because they are independent of $k$. They are purely
classical quantities.

The information about the behavior of the form-factors near the Heisenberg
length is hidden in their off-diagonal parts. The off-diagonal parts of the
form-factors are not independent; the Fourier transform of
Eq.~(\ref{eq:pnt.14}) with respect to $y$ gives us
\begin{eqnarray}
   K_{\text{off}}(x,k) &=& {1\over 4\pi^2\gamma^2}
   \int_{0}^\infty
      \Bigl[
         K^\zeta_{\text{diag}}(x')
         K^\zeta_{\text{off}}(x-x',k)
\nonumber\\
       &+&
         K^\zeta_{\text{diag}}(x')
         K^\zeta_{\text{off}}(-x-x',k)
      \Bigr]
   dx'\;,
\label{eq:uff.3}
\end{eqnarray}
where we substituted the mean density of CPOs from Eq.~(\ref{eq:dfz.8b})  and
also used Eqs.~(\ref{eq:dfn.6e}) and (\ref{eq:dfz.8c}).  In the  derivation
of Eq.~(\ref{eq:uff.3}), we have assumed  that the Heisenberg length is so
large  that Eqs.~(\ref{eq:dfn.6e}), (\ref{eq:dfz.8b}), (\ref{eq:dfz.8c}), and
(\ref{eq:pnt.14}) are valid. We also put Eq.~(\ref{eq:uff.3}) in the form
which is symmetric with respect to $x$. Equation~(\ref{eq:uff.3}) shows that
$K_{\text{off}}(x,k)$ is an analytic function near the Heisenberg length
$x\sim 2\pi \bar d(k)$ in agreement with results of
Ref.\onlinecite{Andreev-Altshuler-95}. A more detailed comparison of results
is possible if we express $K^\zeta_{\text{off}}(x,k)$ in terms of
$K^\zeta_{\text{diag}}(x)$.

The functional equation for the spectral zeta function Eq.~(\ref{eq:dfz.2})
results in symmetry properties of the correlation function
Eq.~(\ref{eq:dfz.4a}). The  Fourier transform of this equation gives us
\begin{equation}
   K^\zeta(x,k) = K^\zeta(2\pi\bar d(k)-x,k)\;.
\label{eq:uff.4}
\end{equation}
The derivative with respect to $x$ of both sides of Eq.~(\ref{eq:uff.4})
can be written as
\begin{equation}
   K^{\zeta\,\prime}_{\text{diag}}(x) +
   K^{\zeta\,\prime}_{\text{off}}(x)
   =
   - K^{\zeta\,\prime}_{\text{diag}}(2\pi\bar d-x)
   - K^{\zeta\,\prime}_{\text{off}}(2\pi\bar d-x)\;,
\label{eq:uff.5}
\end{equation}
where the argument $k$ is omitted at $\bar d$ and at
$K^{\zeta\,\prime}_{\text{off}}(x)$. Near the Heisenberg length
$K^{\zeta}_{\text{diag}}(x)$ is almost constant and its derivative can be
neglected. The same thing is true for  $K^{\zeta}_{\text{off}}(x)$ for small $x$.
We, therefore, remain with
\begin{equation}
   K^{\zeta\,\prime}_{\text{off}}(x,k)
   =
   - K^{\zeta\,\prime}_{\text{diag}}(2\pi\bar d(k)-x)\;.
\label{eq:uff.6}
\end{equation}
This equation together with Eq.~(\ref{eq:uff.3}) expresses
$K_{\text{off}}(x,k)$ in the terms of $K_{\text{diag}}(x)$,
$K^{\zeta}_{\text{diag}}(x)$, and $\bar d(k)$. Equations~(\ref{eq:uff.3}) and
(\ref{eq:uff.6}) solve the problem of the periodic orbit computation of
$K(x,k)$ near $x\sim2\pi\bar d(k)$.

One can understand the important role of the short composite orbits by looking
at another form of Eq.~(\ref{eq:uff.3}) which  can be obtained from
Eq.~(\ref{eq:pnt.12}) by making use of Eqs.~(\ref{eq:pnt.1}),
(\ref{eq:dfn.6e}) and (\ref{eq:dfz.8c}):
\begin{equation}
    K_{\text{off}}(x,k) =
    {1 \over 4\pi\gamma^2}\sum_c e^{-\lambda_c l_c}
    K^\zeta_{\text{off}}(x-l_c,k)\;,
\label{eq:uff.2b}
\end{equation}
valid for $x>0$.
Substitution of the integrated Eq.~(\ref{eq:uff.6}) gives
\begin{equation}
    K_{\text{off}}(2\pi\bar d - x) =\sum_c
    {e^{-\lambda_c l_c}\over 4\pi\gamma^2} 
    \Bigl[
        K^\zeta_{\text{diag}}(x+l_c)-\gamma
    \Bigr]\;.
\label{eq:uff.2}
\end{equation}
The function $K_{\text{off}}(2\pi\bar d - x)$ has spikes for the
small values of $x$ equal to the length differences of the short CPOs.

The inverse Fourier transform of Eq.~(\ref{eq:uff.3}) would give us the
expression for $R_{\text{off}}(\varepsilon,k)$. However, it does not
converge, because $K^{\zeta}_{\text{off}}(x,k)$ remains constant when $x$
goes to infinity. The regularization of the inverse Fourier transforms
obtained  by taking derivatives of Eq.~(\ref{eq:uff.3}) with respect  to $x$
and $2\pi\bar d$ gives
\begin{equation}
   R_{\text{off}}(\varepsilon,k)
   =
   { \cos(  2\pi\bar d(k)\varepsilon ) \over 2\pi^2 \gamma^2}
   \;
   | C_{\text{diag}}(\varepsilon) |^2
   + \bar d(k) \delta(\varepsilon)\;.
\label{eq:uff.7}
\end{equation}
This equation, which reproduces the results of 
Ref.\onlinecite{Andreev-Altshuler-95}, was adopted for the ballistic systems
in Ref.\onlinecite{AAA-dec95} and rederived in
Ref.\onlinecite{Bogomolny-aug96}, see Sec.~\ref{sec:disc} of present work for
more detailed discussion.

It is also instructive to check all of our equations for the case  when the
system exhibits universal behavior. The random matrix theory predicted the
correlation functions of the density of states\cite{Mehta-RMTbook} and the
spectral determinant\cite{Haake-jan96} to be
\begin{eqnarray}
   R(\varepsilon,k) &=&
      { \cos(  2\pi\bar d(k)\varepsilon ) -1 \over 2\pi^2 \varepsilon^2 }
   + \bar d(k) \delta(\varepsilon)\;.
\label{eq:disc.1}
\\
   C(\varepsilon, k) &=& 2\gamma\,e^{i\pi  \bar d(k)\varepsilon}\,
   {\sin(\pi  \bar d(k)\varepsilon) \over \varepsilon}\;,
\label{eq:disc.2}
\end{eqnarray}
where the latter satisfies the symmetry relation Eq.~(\ref{eq:dfz.4a}). The
commonly accepted form of Eq.~(\ref{eq:disc.1}) can be obtained for
$R(\varepsilon,k)/[\bar d(k)]^2$ and it is $\sin^2(z)/z^2$, where $z\equiv
\pi \varepsilon \bar d(k)$.  The universal form-factors are
\begin{eqnarray}
   K(x,k) &=& {|x| + 2\pi \bar d(k) - \left||x|-2\pi \bar d(k)\right|
   \over 8\pi^2}
\label{eq:disc.3}
\\
   K^\zeta(x,k) &=& \gamma\theta(x)\theta(2\pi \bar d(k) - x)\;.
\label{eq:disc.4}
\end{eqnarray}
One can note the symmetry of the universal form-factor Eq.~(\ref{eq:disc.3})
with respect to the exchange $x\leftrightarrow 2\pi\bar d(k)$.

In order to see some kind the of universal length spectrum correlations, let
us use the inverse density of states $k_x$, which is a purely classical
function defined in Eq.~(\ref{eq:disc.7}).  The Fourier transforms of
Eqs.~(\ref{eq:disc.3}) and (\ref{eq:disc.4}) with respect to $k$ and the
change of sign give
\begin{eqnarray}
   \tilde R(x,y) &=& \tilde d_{\text{ppo}}(x)
   {1\over x} \int_0^x dx' \,{\sin(k_{x'}y) \over \pi y}
\label{eq:disc.8}
\\
   \tilde C(x,y) &=& \tilde d_{\text{cpo}}(x)\,
   { \sin( k_{x}y) \over \pi y}\;.
\label{eq:disc.9}
\end{eqnarray}
Both correlation functions satisfy the normalization conditions
Eqs.~(\ref{eq:dfn.5b}) and (\ref{eq:dfz.7a}). Equation~(\ref{eq:disc.9}) is
the ``universal'' correlation function of the composite actions.
Equation~(\ref{eq:disc.8}) is the essence of the relation
Eq.~(\ref{eq:pnt.14}) and it is the most general ``universal'' correlation
function of the actions.


\section{Discussion and Summary}
\label{sec:disc}

The principal results of the present work are the statistical relations
Eq.~(\ref{eq:uff.3}) or Eq.~(\ref{eq:uff.2b}) and the approximate functional
equation Eq.~(\ref{eq:uff.6}). The correlation functions and their
form-factors depend parametrically on the widths of the averaging windows
$\Delta k$ and $\Delta l$. At the same time these parameters do not appear in
the statistical relations Eq.~(\ref{eq:uff.3}) and Eq.~(\ref{eq:uff.2}).
Therefore, these relations are valid for some values of $\Delta k$ and $\Delta
l$ which should be specified.

The probabilistic derivation of Eq.~(\ref{eq:uff.3}) means that
$K_{\text{off}}(x,k)$ and $R(\varepsilon,k)$ possess fluctuations. Therefore,
one has to choose $\Delta k$ and $\Delta l$ in such a way that that these
fluctuations will be smaller than the correlation functions
themselves.\cite{Prange-mar97}

Let us assume that the four-point correlation function of levels has the
universal random matrix theory form that is justified for some chaotic
systems\cite{Bogomolny-feb95,Shukla-apr97}. Then one can show, by making use
of the geometric representation Fig.~\ref{fig:dfn.pol}, that the mean square
fluctuations of $R(\varepsilon,k)$ are of the order of ${\bar d}^{3/2}
\Delta k^{-1/2}$.

Let us assume that the two-point correlation function given by
Eq.~(\ref{eq:uff.7}) behaves like $1/\varepsilon^2$ for  $\varepsilon\bar
d(k)\gg 1$. Then it becomes of the order of $\Delta l^2$ near  $\varepsilon
\sim 1/\Delta l$, which is the maximal value of $\varepsilon$  where the
definition Eq.~(\ref{eq:dfn.3}) is meaningful. Therefore the condition that
the fluctuations of $R(\varepsilon,k)$ are smaller than $R(\varepsilon,k)$
itself is
\begin{equation}
    \Delta l^2 \gg { {\bar d}^{3/2} \over \Delta k^{1/2} }\;.
\label{eq:dfn.10}
\end{equation}
This inequality, together with Eqs.~(\ref{eq:dfn.8}) and (\ref{eq:dfn.9}),
restricts the choice of $\Delta k$ and $\Delta l$ strongly.

The inequalities Eq.~(\ref{eq:dfn.9}) taken near the Heisenberg length become
the approximate equation relating $\Delta l$ to $\Delta k$. We can substitute
$\Delta l(\Delta k)$  into Eq.~(\ref{eq:dfn.10}) and obtain
\begin{equation}
    \Delta k^5 \gg { {\bar d}^3 \over {\bar d}^{\prime 4}}\;.
\label{eq:dfn.11}
\end{equation}
Let ${\cal N}(k)$ be a function counting energy levels with $k_n<k$. One can
apply the approximation ${\bar d'}(k)\approx {\bar d}^2(k) /{\cal N}(k)$ which is valid
for generic system. The counting function can be substituted into
Eq.~(\ref{eq:dfn.11}), and it gives the widths of the averaging window measured
in the number of energy levels
\begin{equation}
    \Delta {\cal N} \gg {\cal N}^{4/5}\;.
\label{eq:dfn.12}
\end{equation}
This inequality has to be satisfied together with $\Delta{\cal N}\ll{\cal N}$.
Therefore the numeric check of relations like
Eqs.~(\ref{eq:uff.2}) and (\ref{eq:uff.7}) is difficult. One has to take at
least ${\cal N}\sim 10^{10}$ and average over $\Delta{\cal N}\sim 10^9$ levels
in order to see non-universal features predicted by Eqs.~(\ref{eq:uff.2}) and
(\ref{eq:uff.7}). The two-point correlation function has to be evaluated for
$\varepsilon$ in the range from zero to $\bar d /\Delta l \sim {\cal N}/\Delta
{\cal N} \sim 10$ level spacings. This estimate also shows that
\begin{equation}
   \Delta k \Delta l \sim {\Delta {\cal N}^2 \over \cal N} \;\gg\;1
\end{equation}
and the definitions of the correlation functions Eqs.~(\ref{eq:dfn.3}) and
(\ref{eq:dfz.4}) are justified.

The basic object characterizing a chaotic dynamical system is the set of mixing
rates. They show how fast the density-density correlations decay. These rates
are zeros of the so-called classical zeta function, see e.g.
Ref\onlinecite{Cvitanovic-90}, which is approximately $C_{\text{diag}}(is)$,
see Ref.\onlinecite{Bogomolny-aug96}.  The functional equation in the form
Eq.~(\ref{eq:uff.6}) implies
\begin{equation}
   \varepsilon C_{\text{off}}(\varepsilon, k) =
   e^{-2\pi i \bar d(k) \varepsilon }\,
   \bigl[ \varepsilon C_{\text{diag}}(\varepsilon) \bigr]^\ast\;,
\label{eq:disc.10}
\end{equation}
where $\varepsilon C_{\text{diag/off}}$ is defined as the inverse Fourier
transform of $K^{\zeta\prime}_{\text{diag/off}}$. Therefore
$C_{\text{off}}(\varepsilon, k)$ is the classical zeta-function modulated by
quantum oscillations. This function contains the information not only on the
mixing properties of the billiard, but also on quantum recurrence.

The association of $C_{\text{diag}}(is)$ with the classical or Ruelle zeta
function is valid in the same approximation as Eq.~(\ref{eq:uff.7}) was
derived. Therefore, Eq.~(\ref{eq:uff.7}) reproduces exactly results of
Ref.\onlinecite{AAA-dec95}, for the case of the system with broken time
reversal symmetry. The result of Ref.\onlinecite{Bogomolny-aug96} goes beyond
the range  of lengths given by Eq.~(\ref{eq:dfn.6f}). One has to make the
replacement in Eq.~(\ref{eq:uff.7})
\begin{equation}
   |C_{\text{diag}}(\varepsilon)|^2
   \;\;\rightarrow\;\;
   [2\pi \bar d(k) \gamma]^2
   \left\langle
     {
        \zeta(k+\varepsilon/2) \, \zeta^\ast(k-\varepsilon/2)
        \over
        \zeta^\ast(k+\varepsilon/2) \, \zeta(k-\varepsilon/2)
      }
    \right\rangle_{\text{diag}} 
\label{eq:disc.11}
\end{equation}
and use the product Eq.~(\ref{eq:dfz.3}) truncated near the Heisenberg length
$l_p\le2\pi\bar d(k)$ in order to obtain the more accurate results of
Ref.\onlinecite{Bogomolny-aug96}.   We performed the more accurate computation
for the prime numbers, see Appendix~\ref{sec:RZexample}, and obtained the
precise form of the Hardy -- Littlewood conjecture, which is equivalent to the
replacement Eq.~(\ref{eq:disc.11}) in Eq.~(\ref{eq:uff.7}).

All calculations in this work were done for chaotic billiards, where the action
of orbit $p$ has the very simple form $S_p=\hbar k l_p$. The present theory can
be generalized for other types of dynamical chaotic systems. The only problem
is that  the correlation function of actions
$\bigl\langle\delta(x-S_p)\delta(x'-S_{p'})\bigr\rangle$ has to be defined by
averaging over the constant mean period ${\partial S_p /\partial E} +
{\partial S_{p'} /\partial E}$. The justification of such procedure was
discussed in Refs.\onlinecite{Argaman-dec93,Doron-prl97}.

The result of the present work cannot be applied to the systems which possess
time reversal symmetry. In this case we should compute $\tilde C_2(X;X')$
or $\tilde R_2(X;X')$ for $\mu\ne \mu'$, i.e. the action correlations of
orbits with different winding numbers. However, the results presented here 
can be generalized for a system having discrete
symmetries\cite{Cohen-Primack-Smilansky}, see also a discussion of this problem
by Leyvraz and Seligman\cite{Leyvraz-Seligman} and Agam {\em et
al}\cite{AASA-97}. In the case of the discrete symmetry of the system, one
should compute the multiplicities of the periodic orbits and correlations of
these multiplicities in order to obtain the spectral statistics.
Probabilistic number theory methods can be used again for this purpose as in
the case of the modular group.\cite{Bogomolny-Leyvraz-Schmit}

In summary  we have shown that the functional equation for the spectral
determinant implies the action correlations.  Quantum-classical time scale
separation for systems with broken time reversal symmetry allows us to
compute all spectral and action correlation functions beyond their universal
random-matrix theory shapes.  One of the central technical points in the
present work is the derivation of the Hardy -- Littlewood conjecture of
prime-number correlations in such a way that it can be used for actions of a
dynamical system.

\acknowledgments

It is my pleasure to thank U.~Smilansky, D.~Cohen,  Z.~Rudnik, H.~Primack,
E.~Bogomolny, IU.~M.~Galperin,  M.~Berry and J.~P.~Keating for very useful
discussions.  This work was supported by Israel Science Foundation.

\appendix


\section{ Normalization of the correlation functions. }
\label{sec:Norm}

The normalization conditions Eqs.~(\ref{eq:dfn.5b}) and (\ref{eq:dfz.7a}) are
generic for the correlation function and we expect that $\sum_{c'} \Bigl[
P(c)P(c')-P(c,c') \Bigr] = P(c)$ and $\sum_{c'}\Bigl[ P(c''\in c)P(c''\in c')
- P(c''\in c\,\cap\,c''\in c') \Bigr] = P(c''\in c)$. The sum over $c'$ in
Eq.~(\ref{eq:pnt.8}) leads to
\begin{equation}
   P(c)
   \approx
   \sum_{c''}
   P(c''\in c)
   \prod_{p\notin c'',\, l_p < l_{c}^\ast}
   \Bigl[
      1 - 2 P(p\in c)
   \Bigr]\;.
\label{eq:Norm.1}
\end{equation}
We can represent $P(c''\in c)$ as a product, and then
\begin{eqnarray*}
   P(c)
   &\approx&
   \prod_{ l_p < l_{c}^\ast}
   \Bigl[
      1 - 2 P(p\in c)
   \Bigr]
   \sum_{c''}
   \prod_{p\in c'',\, l_p < l_{c}^\ast}
   { P(p\in c)  \over
      1 - 2 P(p\in c)
   }
\\
   &\approx&
   \prod_{ l_p < l_{c}^\ast}
   \Bigl[
      1 - 2 P(p\in c)
   \Bigr]
   \left[ 1 + { P(p\in c)  \over
      1 - 2 P(p\in c)
   }\right]
\\
   &=&
   \prod_{ l_p < l_{c}^\ast}
   \Bigl[
      1 - P(p\in c)
   \Bigr]
\end{eqnarray*}
in agreement with Eq.~(\ref{eq:pnt.5}).


\section{ Correlations of primes obtained from correlations of integers.}
\label{sec:RZexample}

We can demonstrate the relation between the correlation functions of composite
and primitive actions using the example of integer and prime numbers. Let us
consider the prime number $p$ as a PPO of length $\log(p)$ and the integer
number $n$ as a CPO of length $\log(n)$. This ``CPO'' may contain repetitions
of ``PPO'', but this is not important for large numbers.

Saying that the ``CPO'' $m$ ``is a part of'' ``CPO'' $n$ means that $m$ is a
divisor of $n$, and we will write $m\divisor n$. This notation allows us to
rewrite Eq.~(\ref{eq:pnt.8}) as
\begin{eqnarray}
   \lefteqn{ P(n)P(n')  -  P(n,n') }
\nonumber\\
   &=&
   \sum_{m}
   \biggl\{
      \Bigl[
         P(m\divisor n) P(m\divisor n') -
         P(m\divisor n \,\cap\,m\divisor n')
      \Bigr]
\nonumber\\
   &\times&
   \prod_{p\notdiv m}
     \Bigl[
        1 - P(p\divisor n) - P(p\divisor n')
     \Bigr]
   \biggr\}\;,
\label{eq:RZ.1}
\end{eqnarray}
where $m$ runs over integers which do not contain powers of primes. Here
$P(n)$ is the probability to find a prime number, and $P(n,n')$ is the
probability to find pair of primes, see
Refs.\onlinecite{Keating-course92,Kac-book2} for precise definitions.

The derivation of the Hardy -- Littlewood expression is based on the
probability\cite{Keating-course92} that $m$ is a divisor of $n$
\begin{equation}
   P(m\divisor n) = {1\over m}\;.
\label{eq:RZ.2}
\end{equation}
Our derivation is based on the probability to find a common divisor
\begin{equation}
  P(m\divisor n \,\cap\,m\divisor n')
  =
  {1\over m}\sum_{l\ne 0}
  \delta_{n-n'-ml}\;,
\label{eq:RZ.3}
\end{equation}
where the Kroneker $\delta$ symbol in the right hand side is equal to one for
$n=n'-ml$ and equal to zero otherwise.  It carries information about
correlations of integers, because if we found with probability $1/m$ that $n$
is multiple of $m$, then all integers $n' = n+ml$  are also multiples of $m$
with the probability 1.

Substituting the probability Eq.~(\ref{eq:RZ.3}) into Eq.~(\ref{eq:RZ.1}) we
obtain
\begin{eqnarray}
   \lefteqn{ P(n)P(n')  -  P(n,n') }
\nonumber\\
   &=&
   \sum_{m}
   \biggl\{
      \Bigl[
         {1\over m^2} - {1\over m}
         \sum_{l\ne 0}
         \delta_{n,\,n'-ml}
      \Bigr]
   \prod_{p\notdiv m}
     \Bigl[
        1 - {2\over p}
     \Bigr]
   \biggr\}\;,
\label{eq:RZ.4}
\end{eqnarray}
and the right hand side of this equation is zero if $m$ is odd.  Indeed, if
$m$ is odd then $p=2$ is not a divisor of $m$ and $1-2/p$ gives zero. The
density of prime numbers, $P(n) = \prod_p (1-1/p)$, must follow the prime
number theorem and this is the criterion for choice of the upper limit in the
products over primes.

The next step, which is not necessary, is to separate the smooth part from
the part containing correlations
\begin{eqnarray}
   P(n,n')
   =
   \sum_{m\divisor(n-n')}
   {1\over m}
   \prod_{p\notdiv m}
     \Bigl[
        1 - {2\over p}
     \Bigr]
   \;,
\label{eq:RZ.5}
\end{eqnarray}
and this expression is non-zero only if $n-n'$ is even. Only terms with even
$m$ contribute and the term with $m=2$ is always present. We can, therefore,
extract the factor 2 from all terms, and we obtain for even $n-n'$
\begin{eqnarray}
   P(n,n') &=& {1\over2}
   \prod_{p\ge 3}
     \Bigl[
        1 - {2\over p}
     \Bigr]
\nonumber\\
   &+&
   \sum_{m\divisor(n-n')}
   {1\over 2m}
   \prod_{p\notdiv m,\,p\ge3}
     \Bigl[
        1 - {2\over p}
     \Bigr]
   \;,
\label{eq:RZ.6}
\end{eqnarray}
where $m$ is odd and does not contain powers of primes. We can rewrite the
sum as a product
\begin{eqnarray}
   P(n,n') &=& {1\over2}
   \prod_{p\ge 3}
     \Bigl[
        1 - {2\over p}
     \Bigr]
   \prod_{p\divisor(n-n'),\,p\ge3}
      { p-1 \over p-2 }
   \;.
\label{eq:RZ.7}
\end{eqnarray}
This product contains the correct enhancement\cite{Cherwell-46} of the
probability to find a pair of primes $n,n'$ by the factor $(p-1)/(p-2)$ per
each prime, which is divisor of $n-n'$.

A similar computation shows that
\begin{equation}
   \sum_{m}
         {1\over m^2}
   \prod_{p\notdiv m}
     \Bigl(
        1 - {2\over p}
     \Bigr)
    = \prod_p
     \Bigl(
        1 - {1\over p}
     \Bigr)^2 = P(n) P(n')\;.
\label{eq:RZ.7a}
\end{equation}
Therefore we have shown that Eq.~(\ref{eq:RZ.5}) follows
Eq.~(\ref{eq:RZ.4}).

The product Eq.~(\ref{eq:RZ.7}) diverges to zero for large $n$. For this
reason the probability $P(n,n')$ is divided by $P(n)=\prod_p (1-1/p)$,
see Ref.~\onlinecite{Hardy-Wright-38}, Chap.~Postscript on prime-pairs.
\begin{equation}
   { P(n,n') \over P(n)P(n')} =
   2\prod_{p\ge 3}
        { 1 - 2/p \over (1-1/p)^2 }
   \prod_{p\divisor(n-n')}
     \Bigl[
        1 + { 1/p \over 1 - 2/p }
     \Bigr]
   \;.
\label{eq:RZ.8}
\end{equation}
This correlation function contains only the convergent products and it allows
one to compute statistics of the zeros of Riemann zeta function with great
accuracy.\cite{Bogomolny-aug96}


\section{ Smoothed correlation function of primes.}
\label{sec:RZsmoothed}

The smoothed form of the correlation function of prime numbers is valid for
$|n-n'|\gg 1$, and it can be obtained directly from Eq.~(\ref{eq:pnt.11})
\begin{equation}
   1- { P(n,n') \over P(n)P(n')}
   \approx
      \sum_{m}\left[ {1\over m^2} -
            P(m\divisor n \,\cap\,m\divisor n')\right]\;.
\label{eq:RZ.9}
\end{equation}
In this equation we used the probabilities of finding the prime number $P(n)$
and the probability of finding the pair of prime numbers $P(n,n')$, see
Refs.\onlinecite{Keating-course92,Kac-book2}, and we substituted $1/m^2$
instead of $P(m\divisor n) P(m\divisor n')$.

The smoothing of the probability to have a common divisor can be done in a
number of different ways. We suggest that the simplest one is
\begin{equation}
   {1\over m} \int_{-m/2}^{m/2} P(m\divisor n \,\cap\,m\divisor n') dn'
   = {1\over m^2} \theta(|n-n'| - {m\over2})\;.
\label{eq:RZ.10}
\end{equation}
The summation over $m$ in Eq.~(\ref{eq:RZ.9}) can be replaced by the
integration and we obtain
\begin{eqnarray}
   { P(n,n') \over P(n)P(n')}
      &\approx&
   1 - \int^\infty\,dm\,{1-\theta(|n-n'| - {m/2})\over m^2}
\nonumber\\
      &\approx&
   1 - {1\over 2|n-n'|}\;.
\label{eq:RZ.11}
\end{eqnarray}
The choice of the lower limit in the integral is not important, because the
integration is performed from $2|n-n'|$. This result is the leading order
expansion of the correlation function in $|n-n'|^{-1}$ and it
coincides with Keating's result.\cite{Keating-course92}

The probability of finding a prime number can be computed from
Eq.~(\ref{eq:RZ.11}) by application of the normalization condition
Eq.~(\ref{eq:dfn.5b})
\begin{eqnarray}
   {1 \over P(n)} &=& \sum_{n'} 
   \left\{  1- { P(n,n') \over P(n)P(n')} \right\}
\nonumber\\
   &\approx& \sum_{|n-n'|<n} {1\over 2|n-n'|} \approx \log(n)
\label{eq:RZ.12}
\end{eqnarray}
according to the prime number theorem\cite{Edwards-book74,Hardy-Wright-38}.


\begin{references}
\bibitem{Ergodic-time-note}
We assume that the density -- density correlations decay as
  $e^{-t/\tau_{\text{erg}}}$ or faster. For simple chaotic billiards this time
  is approximately twice inverse Lyapunov exponent, for more complicated
  systems it is the time of the diffusion through the system.,   .
\end{references}
\end{document}